\documentclass[reprint]{revtex4-2}

\usepackage{graphicx}
\usepackage{dcolumn}
\usepackage{bm,float,physics,bbold,amssymb}
\usepackage{hyperref}
\hypersetup{
  colorlinks   = true,    
  urlcolor     = blue,   
  linkcolor    = blue,   
  citecolor    = blue 
}
\usepackage[mathlines]{lineno}
\newcommand{\mean}[1]{\left \langle #1 \right \rangle}

\begin{document}

\title{Circulation by microwave-induced vortex transport for signal isolation}

\author{Brittany Richman$^{1,2}$}
\email{brr215@umd.edu}
\author{Jacob M. Taylor$^{1,3,4}$}%
\affiliation{$^1$Joint Quantum Institute (JQI), College Park, Maryland 20742, USA\\
$^2$University of Maryland, College Park, Maryland 20742, USA\\
$^3$Joint Center for Quantum Information and Computer Science (QuICS), College Park, Maryland 20742, USA\\
$^4$National Institute of Standards and Technology (NIST), Gaithersburg, Maryland 20899, USA
}

\date{\today}

\begin{abstract}
Magnetic fields break time-reversal symmetry, which is leveraged in many settings to enable the nonreciprocal behavior of light. This is the core physics of circulators and other elements used in a variety of microwave and optical settings. Commercial circulators in the microwave domain typically use ferromagnetic materials and wave interference, requiring large devices and large fields. However, quantum information devices for sensing and computation require small sizes, lower fields, and better on-chip integration. Equivalences to ferromagnetic order---such as the XY model---can be realized at much lower magnetic fields by using arrays of superconducting islands connected by Josephson junctions. Here we show that the quantum-coherent motion of a single vortex in such an array suffices to induce nonreciprocal behavior, enabling a small-scale, moderate-bandwidth, and low insertion loss circulator at very low magnetic fields and at microwave frequencies relevant for experiments with qubits.

\end{abstract}

\maketitle

\section{\label{intro}Introduction}

In order to scale up quantum computing devices, the supporting hardware for qubits must appropriately scale down. Essential components include elements for signal routing and isolation, such as circulators, as well as low-noise amplifiers, filters, and cooling systems. While development in improving miniaturization and effectiveness of these components progresses apace \cite{devamp,TWPAlincolnLab,devcirc,4modeTA,ETH100,IBMisolator,intelCryoLogic,IBMisolator2020,rosenthal2020}, traditional circulators remain bulky and require high magnetic fields for operation.

Creating a circulator requires nonreciprocal behavior induced by implicit or explicit breaking of time-reversal symmetry. This is then combined with microwave engineering to enable unidirectional transmission with little to no loss in a clockwise or counterclockwise fashion over a large operating bandwidth. The nonreciprocal behavior observed in commercial ferrite junction circulators, which are commonly utilized in smaller systems, is a result of the Faraday effect. The interaction between a magnetic field, e.g., from a permanent magnet, and the central ferrite discs of the device create circulation via wave interference and magnetically induced gyration \cite{pozar}. As a result, commercial ferrite junction circulators are fundamentally limited in size by the signal wavelength. However, their passive operation necessarily eliminates any additional control hardware. Nonetheless, with their large size and reliance on strong fields, the prospect of employing ferrite circulators in large quantum computing schemes is a challenge for the field, though some progress in wavelength-size devices~\cite{SchusterSimon} has been made. This requires finding small-volume, on-chip solutions that operate at low magnetic field and enable the isolation of quantum devices and corresponding precise qubit control and measurement. 

One partial solution to this problem is generating nonreciprocity via the quantum Hall effect, where large magnetic fields break time-reversal symmetry and quantum Hall edge modes are utilized in passive devices to observe circulation \cite{hallvg,mahoney,*Mahoney2017}. Other proposed solutions are actively controlled and take a nonmagnetic or small-field approach, using frequency conversion in driven systems and irreversible dynamics (loss) or synthetic gauge fields in a Floquet basis to generate nonreciprocity \cite{devcirc,4modeTA,BS,KJ,chapman,IBMgyrator,chapman2019}. However, due to their active nature, they necessarily require additional control hardware that is undesirable in larger-scale systems. To date, the only passive, low-field, subwavelength approach uses particle tunneling in small Josephson junction arrays (JJAs) or quantum phase slip junctions, utilizing their behavior in magnetic fields---the Peierls phase in tunneling---to break time-reversal symmetry \cite{koch, muller}. Both of these proposed devices suffer from challenges in implementation due to susceptibility to charge noise, difficulties in implementing robust quantum phase slip devices, and poor performance in operational bandwidth.

\begin{figure}
\centering
\includegraphics[width=0.48 \textwidth]{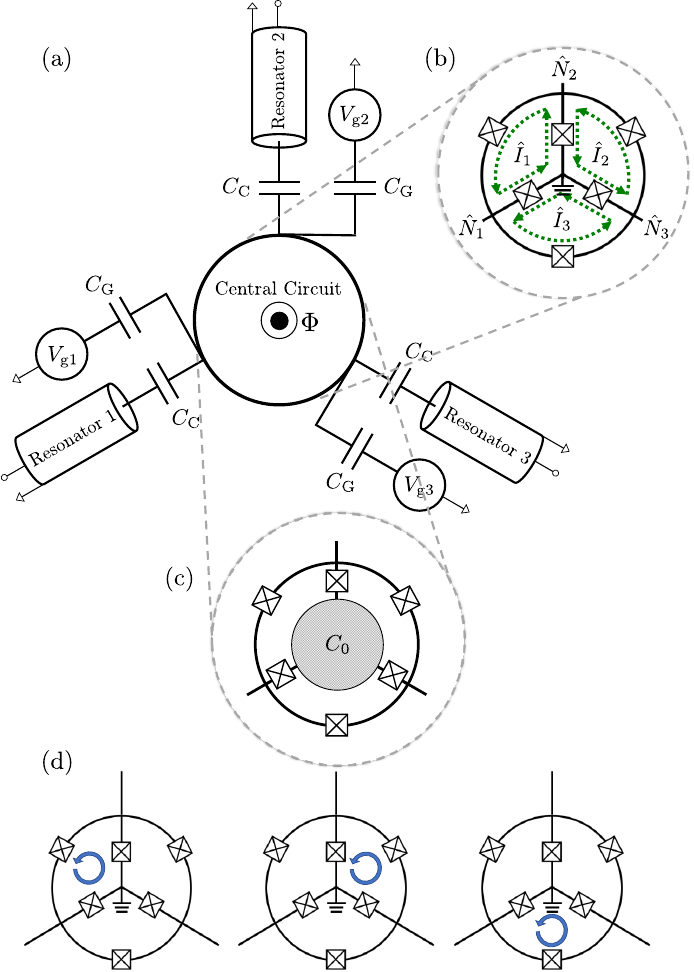}
\caption{(a) A schematic representation of the model with a generic central circuit threaded by external flux $\Phi$ via an external magnetic field applied perpendicular to the plane. Each resonator (enumerated above) and voltage source $V_{\text{g}i}$ is coupled to an island of the central circuit via capacitors $C_\text{C}$ and $C_\text{G}$, respectively. (b) The central circuit model, with identical Josephson junctions $\boxtimes$ having Josephson energy $E_\text{J}$ and capacitance $C_\text{J}$, with labeled loop currents through each third of the model: $\hat{I}_1$, $\hat{I}_2$, and $\hat{I}_3$. (c) An alternative circuit design (explored in Appendix \ref{app:float}) that replaces the central ground with a large central island with self-capacitance $C_0$. (d) A visual of the persistent-current vortices that may exist in each third of the circuit.}
\label{fig:schematic}
\end{figure}

Here we address these challenges by moving from a Cooper pair tunneling device to a persistent current (vortex) tunneling device via insertion of additional Josephson junctions. Our approach, shown schematically in Figs.~\ref{fig:schematic}a and \ref{fig:schematic}b, builds upon the three-island Josephson junction loop proposed by Koch \textit{et al.} \cite{koch} and further explored by M$\ddot{\text{u}}$ller \textit{et al.} \cite{muller}. In contrast to those works, we suggest operating in a non-charge-conserved, intermediate regime of Josephson energy and charging energy being similar. By keeping the Josephson energy similar to the charging energy, we can maintain quantum-coherent vortex dynamics rather than having the microwave response behavior be dominated by spin waves, as occurs at large Josephson energy. We show a reduction in sensitivity to charge noise while enabling large bandwidth operation. 

To understand this performance, we draw an analogy to bulk superconductors, where the Meissner effect leads to persistent currents around holes in a superconductor. In the limit of large tunneling rates, these currents are fixed to each hole. This is related to the dual picture of a lattice of vortices with large effective mass trapped around each hole. However, as the kinetic term increases by decreasing capacitance, and phase-slips are made available by the introduction of tunnel junctions (Josephson junctions), the vortices are able to move. Some devices, such as the flux qubit \cite{FQ1,FQ2}, have demonstrated the ability to prepare a superposition of different vortex states for a small circuit \cite{VS1,VS2}. We suggest that the successful operation of our proposed device arises by exploiting the behavior of such coherent vortices in a magnetic field, analogous to the behavior of the ferrite system where the XY model of the junctions acts as an effective ferromagnet. Here, we propose circulation is achieved via the rotation of these persistent current vortices: the voltage induced by an incoming signal generates a vortex in the circuit, this vortex then rotates about the circuit, leading re-radiation into each port.

This paper is organized as follows. Section~\ref{vortexthings} introduces the central circuit model to be explored, motivating this choice by presenting an investigation of vortices in the model and their potential role in device functionality. Section~\ref{S1} couples the central circuit to external circuits to investigate transmission properties of the system. A theoretical analysis in the single excitation limit is presented, resulting in the single photon scattering matrix (S-matrix), followed by conditions for ideal circulation. Numerical results related to bandwidth, circulation, and noise performance are shown in Section~\ref{S5}, and we conclude and provide an overview of results in Section~\ref{conclusion}. In the appendices that follow, we explore an alternative circuit model to the one presented in Section~\ref{vortexthings}, followed by a numerical examination of two flux points in addition to the one investigated in Section~\ref{S5}. We then discuss various parameter effects and provide additional details to supplement the noise analysis presented in Section~\ref{S5}.

\section{\label{vortexthings} Vortex-based nonreciprocity}

\subsection{Model circuit}
\label{ccmod}

Here we investigate a minimal model of a circuit for our proposed system which consists of three superconducting islands connected to each other and a central grounded island via Josephson junctions, as shown in Fig.~\ref{fig:schematic}b, all threaded by an external flux $\Phi$. We note that the central ground can, in principle, be replaced by a capacitive island with $C_0 \gg C_\text{J}$, the nominal capacitance of each island. This alternative circuit is shown in Fig.~\ref{fig:schematic}c and explored in detail in Appendix \ref{app:float}. For the circuit in Fig.~\ref{fig:schematic}b, the overall Hamiltonian is:
\begin{equation}
\label{eq:cc}
    H_{\text{CC}}=4 E_{\Sigma} \ ( \underline{\hat{N}}-\underline{N}_\text{g})^T \underline{\underline{K}}^{-1} (\underline{\hat{N}}-\underline{N}_\text{g}) + E_\text{J} \ V(\underline{\hat{\varphi}},\Phi) \enspace .
\end{equation}
The operators $\underline{\hat{N}}$ and $\underline{\hat{\varphi}}$ are the vectors of number and phase operators for each island of the circuit, with individual elements satisfying the commutation relation $[\hat{\varphi}_j,\hat{N}_k]= i \delta_{jk}$. $E_\text{J}$ is the Josephson energy of each junction and $E_{\Sigma}$ is the circuit's charging energy, defined as $E_{\Sigma} = \frac{e^2}{2 C_\text{M}}$ where $C_\text{M}$ is the largest eigenvalue of the circuit's capacitance matrix $\underline{\underline{C}}$, defined in Eq.~(\ref{eq:cmat}). $\underline{\underline{K}}^{-1}$ is the dimensionless inverse capacitance matrix appropriate for the circuit, found by factoring $C_\text{M}$ from $\underline{\underline{C}}^{-1}$. As it is written, $H_{\text{CC}}$ only stipulates the central circuit be a superconducting, three-island system containing identical Josephson junctions. By specifying the elements of $\underline{\underline{K}}^{-1}$ and the tunneling potential $V(\underline{\hat{\varphi}},\Phi)$ we establish the particular central circuit for the system explored in this work. This allows us to compare to prior work without substantial computational difficulty.

We define $\underline{\underline{K}}^{-1}$ by inverting the capacitance matrix $\underline{\underline{C}}$,
\begin{equation}
\label{eq:cmat}
    \underline{\underline{C}}=  \begin{bmatrix}
    C_\text{S} &-C_\text{J}&-C_\text{J}\\
    -C_\text{J}& C_\text{S}&-C_\text{J}\\
    -C_\text{J}&-C_\text{J}& C_\text{S}
    \end{bmatrix} \enspace ,
\end{equation}
where $C_\text{S} = C_\text{C} + C_\text{G} + 3C_\text{J}$ with $C_\text{J}$ the capacitance of each Josephson junction, $C_\text{C}$ the coupling capacitance, and $C_\text{G}$ the residual capacitance to ground. For this matrix, $C_\text{M} = C_\text{C}+C_\text{G}+4C_\text{J}$ is the largest eigenvalue. Factoring $C_\text{M}$ from $\underline{\underline{C}}^{-1}$ we can then find:
\begin{equation}
\label{eq:kimat}
\underline{\underline{K}}^{-1} = \begin{bmatrix}
    1+\delta&\delta&\delta\\
    \delta&1+\delta&\delta\\
    \delta&\delta&1+\delta
    \end{bmatrix} \enspace ,
\end{equation}
where $\delta = \frac{C_\text{J}}{C_\text{C}+C_\text{G}+C_\text{J}}$. 

The inductive terms, corresponding to tunneling through the Josephson junctions, are described by
\begin{equation}
\label{eq:tp}
\begin{split}
V(\underline{\hat{\varphi}},\Phi) \ = \ - \sum_{i=1}^3 [ \cos&(\hat{\varphi}_{i+1}-\hat{\varphi}_i- 2 \pi A) \\
 + &\ \cos(\hat{\varphi}_i) ]
\end{split} \enspace ,
\end{equation}
where we take $i+1=4$ and $i=1$ to be the same. We note that this potential is equivalent to the XY model where the phase $\varphi$ represents the angle of the $U(1)$ degrees of freedom.

The acquired Peierls phase factor, $2 \pi A$, is equal to the line integral of the vector potential in a counterclockwise fashion around each third of the device:
\begin{equation}
\label{eq:Adef}
    2 \pi A = \frac{2 \pi}{\Phi_{\text{0}}} \oint \mathbf{A} \cdot d\bold{l} = \frac{2\pi\Phi}{3\Phi_{\text{0}}}\ .
\end{equation}
This $A$ corresponds to the frustration of the circuit, i.e., the flux penetration per loop of the device in terms of the magnetic flux quantum. 

In what follows, we denote the eigenstates of $H_{\text{CC}}$ as $\ket{\varepsilon_m}$ with eigenenergy $\varepsilon_m$, and label the ground state $\ket{G}$. These are parametric functions of $A$, and periodic in $A$.

\subsection{Entering the vortex regime}
\label{VT1}

The behavior and properties of the central circuit, shown in Fig.~\ref{fig:schematic}b, depend on the relative size of the two characteristic energy scales in the system: the energy associated with Cooper pair tunneling between islands, $E_{\text{J}}$, and the energy associated with putting a single electron on an island, $E_\text{C} \sim E_{\Sigma}$. In the two extreme energy limits, $E_{\text{J}} \ll E_\text{C}$ or $E_{\text{J}} \gg E_\text{C}$, the relevant excitations are charges or vortices, respectively \cite{TF,fazioreport,ballisticvortices}. Note that charges have the same properties in the charged-dominated regime that vortices have in the tunneling-dominated regime \cite{fazio,fazioreport}. In the charge-dominated regime, where $E_{\text{J}} \ll E_\text{C}$, charges are well-defined, massive particles that feel a Lorentz force in the presence of an electromagnetic field and acquire an Aharanov-Bohm phase (represented in the charge basis as a Peierls phase) when moving around an area containing a magnetic field~\cite{griffiths,sakurai}. However, in this regime, external voltages and other sources of charge disorder, such as offset charge, have a dramatic effect on circuit behavior due to the very small capacitances necessary to achieve large $E_\text{C}$.

In the classical tunneling limit, where $E_{\text{J}} \gg E_\text{C}$, vortices are the topological excitations of the XY model that behave as massive (and essentially classical) particles. They respond to the presence of external current by experiencing a force perpendicular to the current flow (a Lorentz force), and the presence of external charge on the islands affects a moving vortex the same way a magnetic field affects a moving charge (an Aharanov-Casher effect for vortices) \cite{fazio,fazioreport,ballisticvortices,lobb}. In this way, charges and vortices are dual to each other, with these mirrored properties being manifestations of the charge-vortex duality that exists in JJAs. However, the vortex domain has two characteristic excitations: the topological excitations, as just described, as well as plasma oscillations (spin waves in the XY model language) that correspond to the response of a classical circuit of inductors and capacitors with values set by the expectation value of the Josephson inductance for the nominal ground state (see, e.g., Ref.~\cite{cosmic}). The plasma oscillations do not break time-reversal symmetry; in essence, the Meissner effect traps flux in the vortices, and thus, all time-reversal symmetry-breaking terms arise only from vortex motion. When $E_\text{J} \gg E_\text{C}$, this motion becomes exponentially slow due to the challenge of tunneling between allowed current configurations, and thus, one cannot make a circulator in this regime.

We wish to identify an energy regime using this duality, where time-reversal symmetry is broken in the microwave regime and charge noise is not relevant. We thus focus on $E_\text{J} \sim E_\text{C}$ as a regime where vortices can move and charge noise may be reduced, enabling the potential for noise-resistant nonreciprocity in the form of unidirectional signal transmission that is robust in the presence of random variation in offset charge. In this way, the model may act as an ideal circulator. 

This intermediate regime has been partially explored by Koch \textit{et al.} \cite{koch} and M$\ddot{\text{u}}$ller \textit{et al.} \cite{muller}. However, in those works, the system was defined with a charge-conserving three-island Josephson junction loop, and total charge on the small capacitance region was fixed. In what follows, we leverage the connection of each island to ground via an additional Josephson junction to enable large charge number fluctuations and suppress charge noise. This leads to circulator performance driven by vortex behavior rather than Cooper pair behavior.

\subsection{Quantum behavior of vortices in the intermediate tunneling regime}
\label{VT2}

To explore the vortex picture, we set $E_\text{J} = 4E_\text{C}$ to ensure that there are sufficient phase fluctuations to warrant a quantum treatment of the vortex dynamics, and numerically diagonalize $H_\text{CC}$ [given by Eq.~(\ref{eq:cc})] for a range of $A$ from 0 to 0.5, as the eigenvalues are symmetric in $A \rightarrow -A$ and periodic (same for $A+1$ as for $A$). We denote the corresponding energy eigenstates $\ket{\varepsilon_n}$ and eigenvalues $\varepsilon_n$, with implicit dependence on $A$ understood. We note for our chosen parameters ($E_\text{J} = 4E_\text{C} \approx 45$ GHz) and eventual optimum operating point (near $A=0.2478$), that the lowest energy spin-wave excitation occurs at $20$ GHz, far higher than the low-energy excitations (in the few GHz regime) we consider here. As a result, we expect the predominant dynamics to be that of vortices, not spin waves.

\begin{center}
\begin{figure}
\includegraphics[width = 0.48 \textwidth]{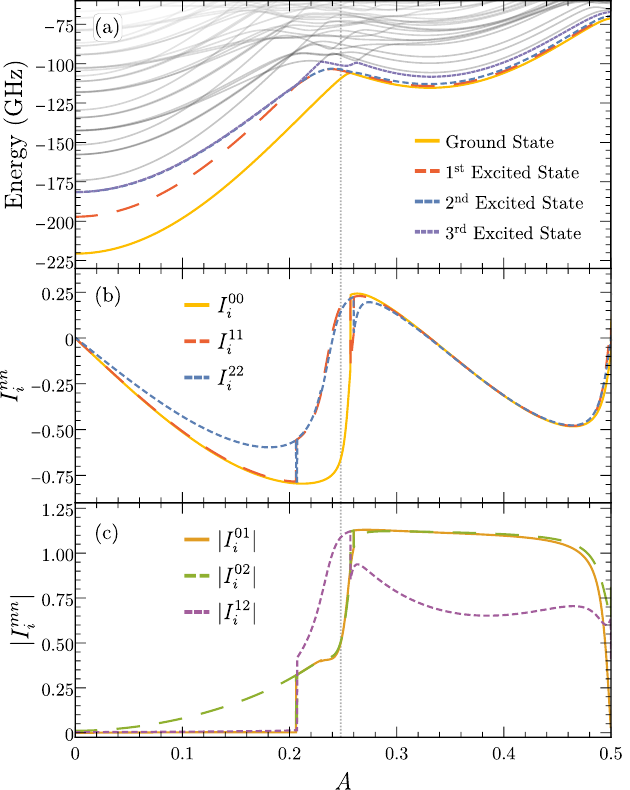}
\caption{(a) The first fifty energy eigenvalues as a function of $A$, the flux penetration per loop. The relevant levels for this analysis are indicated in the legend; higher levels are grayed out. (b) Expectation values of the persistent currents in the ground state, first excited state, and second excited state for each loop ($i=1,2,3$) of the circuit shown in Fig.~\ref{fig:schematic}b as a function of $A$. (c) Magnitudes of three of the off-diagonal elements of the loop-current operators in the eigenbasis for each loop ($i=1,2,3$) as a function of $A$. Note that the true currents require re-scaling by $I_{\text{C}}$ (here taken to be one). The circuit is periodic on $A\in \lbrace 0, 1 \rbrace$ and anti-symmetric about $A=0.5$, hence only $0 \leq A \leq 0.5$ is shown here. Parameters chosen for these plots are $E_\text{J}/h = 4 E_\text{C}/h \approx 45 \text{ GHz}$, where $E_\text{C} = \frac{e^2}{2C_\text{J}}$, $E_\text{J}/E_{\Sigma} \approx 40$, $C_\text{C} = 6 C_\text{J}$ and $C_\text{G} = 0.02 C_\text{J}$. Offset charge is tuned such that the total Cooper pair number in the ground state is $\approx$ 1.2, with $N_{\text{g}1}=N_{\text{g}2}=N_{\text{g}3}\approx 0.4$.}
\label{fig:ELamps}
\end{figure}
\end{center}

We approximate the full Hilbert space using a charge basis $\ket{n_1,n_2,n_3}$ for Cooper pairs on each island, with $|n_i|\leq 4$. We checked truncation and found no difference by incrementing the maximum allowed charge by one. This creates a Hilbert space dimension of $9^3$ and is straightforward to solve on a personal computer. The energy eigenvalues of the circuit as a function of $A$ are shown in Fig.~\ref{fig:ELamps}a. We note the presence of level-crossings at $A \approx 0.2$ and $A \approx 0.26$ within the first four energy levels due to the high symmetry of the system. 

We now consider the low-energy space of this system, and examine the presence and behavior of localized vortices. For low-frequency excitations (in the few GHz regime), and starting from the circuit ground state, we can expect to only explore superpositions of the first few eigenstates. To see the vortices, we identify the operators associated with the persistent currents around each of the small loops in our central circuit, shown in Fig.~\ref{fig:schematic}b:
\begin{equation}
\label{eq:lco}
\begin{split}
\hat{I}_i = & \ \sin(\hat{\varphi}_{i+1}-\hat{\varphi}_i-2 \pi A) \\
 &+ \ \sin(\hat{\varphi}_i) \ - \ \sin(\hat{\varphi}_{i+1})  \enspace ,
\end{split}
\end{equation}
where we take $i+1=4$ and $i=1$ to be the same. To find the true current in the system, one multiplies Eq.~(\ref{eq:lco}) by $I_{\text{C}}$, the single-junction critical current. Fig.~\ref{fig:schematic}d shows a schematic representation of each of these loop currents for our model of interest. 

\begin{figure}
\centering
\includegraphics[width = 0.48 \textwidth]{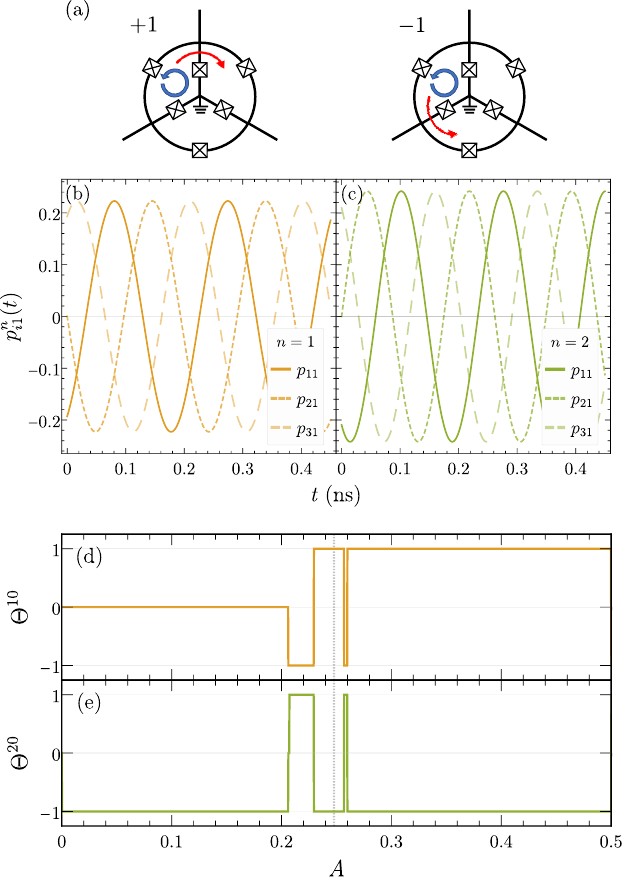}
\caption{In (a), a schematic of the clockwise or counterclockwise vortex tunneling shown in (b)-(e). The linear response of the loop currents in the central circuit due to the presence of a voltage on the first island is shown in (b)-(c). The contributions from the (b) first excited state and (c) second excited state are shown, demonstrating the creation and motion of vortices as a result of an incoming signal at our chosen operating point $A \approx 0.2478$. Vortex tunneling is shown in the two-state superpositions of (d) the ground state and first excited state and (e) the ground state and second excited state as a function of $A$, the flux penetration per loop. As in Fig.~\ref{fig:ELamps}, the circuit is periodic on $A\in \lbrace 0, 1 \rbrace$ and anti-symmetric about $A=0.5$, hence only $0 \leq A \leq 0.5$ is shown here. +1.0 indicates clockwise vortex tunneling, -1.0 indicates counterclockwise vortex tunneling, and 0 indicates the absence of a vortex. Parameters for these plots are the same as Fig.~\ref{fig:ELamps}.}
\label{fig:directions}
\end{figure}

A local vortex is identified by a nonzero loop current in one or two of the three loops. Note that if all three loop currents are equal, there is only a persistent current circulating about the exterior. For the energy eigenstates, Fig.~\ref{fig:ELamps}b shows that the expectation values of each of the three loop currents are the same, and thus the eigenstates do not correspond to localized vortex states, though each eigenstate has a distinct persistent current around the exterior of the circuit. We also see that the level crossings manifest themselves in the subsequent persistent-current analysis, proving them to be regions of flux that create dramatic changes in behavior due to changes in the nature of the ground state, as seen in Figs.~\ref{fig:ELamps}b, \ref{fig:ELamps}c, \ref{fig:directions}d, and~\ref{fig:directions}e. Our chosen operating point for circulation, shown with a vertical grey line, is away from these crossings.

While there are no localized vortices for energy eigenstates, an incoming microwave photon can lead to the excitation of superpositions of circuit states. To examine this, we consider the matrix elements of the three loop currents over the first three eigenstates $\{ \ket{\varepsilon_0} (=\ket{G}),\ket{\varepsilon_1},\ket{\varepsilon_2} \}$, denoted $I^{kl}_i = \bra{\varepsilon_k} \hat{I}_i \ket{\varepsilon_l}$, whose magnitudes are shown in Figs.~\ref{fig:ELamps}b and \ref{fig:ELamps}c. The symmetry of the circuit requires that the absolute value of each of the loop-current operator matrix elements be independent of $i$, that is, the loop to which they correspond. However, as we now show, superpositions of eigenstates exhibit localized vortices due to nontrivial phases of the off-diagonal matrix elements $I^{kl}_i$ for $k\neq l$. 

We write the argument of the $I^{mn}_i$ matrix elements as $\phi_i^{mn}$. The differences $\phi_i^{mn} - \phi_{i+1}^{mn}$ only take the values $\{0,\pm 2\pi/3\}$, as expected by the three-fold symmetry of the system. We now consider their role in circulation. Examining a superposition $c_m \ket{\varepsilon_m} + c_n \ket{\varepsilon_n}$ with $m>n$, we see that the $i^{\text{th}}$ loop-current operator has an expectation value at time $t$ of
\begin{equation}
\label{eq:ETtwo}
\begin{split}
\langle \hat{I}_i (t) \rangle \ = \ &|c_m|^2 I_i^{mm} \ + \ |c_n|^2 I_i^{nn} \\
&+ \ 2 |c_m| |c_n| |I_i^{mn}| \\
& \quad \quad \quad \times \cos \left( \omega_{mn} t - \theta_{mn} + \phi_i^{mn} \right) \enspace ,
\end{split}
\end{equation}
where $\phi_i^{mn}$ is the argument of the off-diagonal matrix element $I_i^{mn}$, $\theta_{mn}$ is the argument of $c_m^* c_n$, and $\omega_{mn} = \frac{(E_m-E_n)}{\hbar}$. Thus, we find a superposition of two states has, in general, a local vortex, and that these local vortices circulate (countercirculate) about the system when $\phi_i^{mn} - \phi_{i+1}^{mn} = \pm 2 \pi/3$ at a frequency $\omega_{mn}$. This is shown schematically in Fig.~\ref{fig:directions}a.

Accordingly, we define the directional function for the difference in $\phi_{i}^{mn}$ variables modulo $2\pi$ as
\begin{equation}
\label{eq:directionality}
\Theta^{mn} = \begin{cases}
    					+1  \\
    					\ \ 0 \\
    					-1 
				\end{cases} \text{if } \phi^{mn}_i-\phi^{mn}_{i+1} = \begin{cases}
																		+2\pi/3\\
																		0 \\
																		-2\pi/3
																	\end{cases} \enspace , 
\end{equation}
to allow us to plot the behavior of potential local vortices for different values of $m,\ n,$ and $A$, including the existence or absence of vortex circulation and what direction. $\Theta^{mn}$ takes the value +1 for clockwise circulation, -1 for counterclockwise circulation, and 0 for no circulation (absence of a local vortex). The directionality of vortex circulation associated with the two-state superpositions $0-1$ and $0-2$ as a function of $A$ are shown in Figs.~\ref{fig:directions}d and \ref{fig:directions}e, respectively.

\subsection{Connecting vortex behavior with the microwave response}
\label{LR}

We now consider what happens if an oscillating voltage is applied to the $j^\text{th}$ island of the central circuit via a coupling capacitor. We denote the coupling as $\hbar \Omega_j(t) \hat{B}_j$, with $\underline{\hat{B}} = \underline{\underline{K}}^{-1} \underline{\hat{N}}$ the vector of dimensionless voltages on each island. Using linear response theory, we can evaluate the expected loop currents at a later time for a perturbation $\Omega_j(t) \propto \delta(t)$ as
\begin{equation}
    P_{ij}(t) = -i \mean{[\hat{I}_i(t),\hat{B}_j(0)]} \enspace ,
\end{equation}
where we use $P$ as in a dimensional setting this quantity has units of power, scaling as $\frac{2 e I_{\text{C}}}{C_{\text{M}}}$, and can be interpreted as the power in the circuit induced by the drive. Using the Green's function relations, we then also expect that a monochromatic excitation at frequency $\omega$ gives the Fourier transform of this value for the expected long-time response of the different loop-current operators. We will also examine, for the microwave response, the voltage-voltage correlation function, 
\begin{equation}
\label{eq:VV}
    T_{ij}(t) = -i \mean{[\hat{B}_i(t),\hat{B}_j(0)]} \enspace ,
\end{equation}
whose Fourier transform will be identified later as the transmission matrix of the circuit.

Working this out explicitly, 
\begin{equation}
\begin{split}
    P_{ij}(t) &= -i \bra{G} [e^{i H t} \hat{I}_i e^{-i H t}, \hat{B}_j] \ket{G} \\
     &= -i \bra{G} \hat{I}_i \sum_n e^{-i \varepsilon_n t} \ketbra{\varepsilon_n}{\varepsilon_n} \hat{B}_j \ket{G} + \textrm{h.c.} \enspace , \label{eq:linresponse} \\
\end{split}
\end{equation}
and
\begin{equation}
    \tilde{P}_{ij}(\omega) = -i \bra{G} \hat{I}_i \sum_n \frac{\ketbra{\varepsilon_n}{\varepsilon_n}}{\omega - \varepsilon_n + i \gamma/2} \hat{B}_j \ket{G} \enspace ,
\end{equation}
where $\gamma \rightarrow 0^+$ is taken to ensure causal ordering and $P_{ij}(\omega) = \tilde{P}_{ij}(\omega) + \tilde{P}^*_{ij}(-\omega)$. We note that a narrowband signal will preferentially excite states near the signal frequency, as is apparent in $\tilde{P}_{ij}(\omega)$'s denominator.

Examining Eq.~(\ref{eq:linresponse}), we can interpret the linear response behavior as the evaluation of $\langle \hat{I}_i \rangle$ for a state of the form $\ket{G} - i  \sum_n \ketbra{\varepsilon_n}{\varepsilon_n} (\sum_j \Omega_j \hat{B}_j) \ket{G} $ for $\Omega_j \rightarrow 0$. Thus we can conclude that, insofar as $\hat{B}_j$ produces superpositions of energy eigenstates, we can expect the creation of moving vortices within the array upon an incoming signal. As we expect sinusoidal signals near the $0-1$ and $0-2$ transition frequencies, we plot the terms of $P_{i1}(t)$ corresponding to the first and second excited states in Figs.~\ref{fig:directions}b and \ref{fig:directions}c [denoting the $n^\text{th}$ term of $P_{i1}(t)$ as $p^n_{i1}(t)$] to illustrate the basic expected dynamics of the system.

Noting that $-P_{1j}(-t)$ corresponds to the expected voltage at island $j$ due to a current at loop $1$, we can see the emergence of circulator dynamics: a voltage creates vortices, which rotate, and re-radiate into the leads. Crucially, we note that the observed rotation of vortices corresponds to highly nonreciprocal behavior, and thus we can expect generically that this central circuit will provide the opportunity for building a circulator. However, substantial practical details must now be worked out. We do so below, focusing on the linear response (low power) regime.

\section{\label{S1}Central circuit coupled to external circuits}

\subsection{The full circuit model}
We now consider the central, three-island superconducting circuit capacitively coupled at each island to a resonator and voltage source, as shown in Fig.~\ref{fig:schematic}a. These resonators are included to enable effective impedance matching between incoming microwave signals and the central circuit, and will be tuned in frequency and coupling to enable optimal performance. Each resonator port is opened to an external source of microwaves, i.e., a transmission line, allowing signals to flow into and out of the system. The central circuit, specified above in Section~\ref{ccmod}, is threaded by an external flux $\Phi$ via the application of an external magnetic field.

The Hamiltonian for this model is composed of terms corresponding to the transmission lines, resonators, and central circuit, as well as each circuit-resonator interaction and resonator-transmission line interaction. Generally, this may be written as:
\begin{equation}
\label{eq:GH}
H \ = \ H_{\text{TL}} \ + \ H_{\text{R}} \ + \ H_{\text{CC}} \ + \ H_{\text{CC-R}} \ + \ H_{\text{R-TL}} \enspace .
\end{equation}
Taking the weak-coupling limit, the rotating-wave approximation, and the Markov approximation, each term of the Hamiltonian in Eq.~(\ref{eq:GH}) can be expressed as:
\begin{equation}
\label{eq:Hterms}
\begin{split}
H_{\text{TL}}&=\sum_{k=1}^3 \int d\nu \ \hbar \nu \  \hat{b}^{\dagger}_k (\nu) \hat{b}_k(\nu) \\
H_{\text{R}}&=\sum_{k=1}^3 \ \hbar \omega_\text{R} \  \hat{a}^{\dagger}_k \hat{a}_k \\
H_{\text{CC}}&=4 E_{\Sigma} \ ( \underline{\hat{N}}-\underline{N}_\text{g})^T \underline{\underline{K}}^{-1} (\underline{\hat{N}}-\underline{N}_\text{g}) \\
& \qquad + E_\text{J} \ V(\underline{\hat{\varphi}},\Phi) \\
H_{\text{CC-R}}&=\sum_{k=1}^3 \ \hbar g \  (\hat{a}_k+\hat{a}^{\dagger}_k) \hat{B}_k\\
H_{\text{R-TL}}&=\sum_{k=1}^3 \int d\nu \ i\hbar\sqrt{\frac{\kappa}{2\pi}} \ \left( \hat{b}^{\dagger}_k(\nu) \hat{a}_{k}-\hat{a}^{\dagger}_{k} \hat{b}_k (\nu) \right)
\end{split} \enspace .
\end{equation}
$H_{\text{TL}}$ \cite{colgard} describes each transmission line as containing an identical continuous spectrum of modes over frequency $\nu$. $H_{\text{R}}$ \cite{koch} is the contribution from the system's three identical resonators, consisting of a single low-lying mode with a characteristic frequency $\omega_\text{R}$. The operators $\hat{a}^{\dagger}_k$ and $\hat{a}_k$ create or destroy a photon in the $k^{\text{th}}$ resonator and satisfy the usual commutation relation: $[\hat{a}_j,\hat{a}^{\dagger}_k]=\delta_{jk}$. Similarly, $\hat{b}^{\dagger}_k(\nu)$ and $\hat{b}_k(\nu)$ create or destroy a photon in the $k^{\text{th}}$ transmission line and satisfy the commutation relation: $[\hat{b}_j(\nu),\hat{b}^{\dagger}_k(\nu')]=\delta_{jk}\delta(\nu-\nu')$.

$H_{\text{CC-R}}$ and $H_{\text{R-TL}}$ \cite{colgard,koch} are the interactions between each circuit island and resonator and between each resonator and transmission line, respectively. By treating all resonators and all transmission lines as identical, we take the coupling constants, $g$ and $\kappa$, to be identical across all three interactions, and will later optimize to ensure the best circulator performance. The operator $\hat{B}_k$, as defined in Section~\ref{LR}, is a dimensionless operator corresponding to the dimensionless voltage on the $k^{\text{th}}$ island, defined as the $k^{\text{th}}$ row of $\underline{\hat{B}} \ = \ \underline{\underline{K}}^{-1} \underline{\hat{N}}$.

Lastly, $H_{\text{CC}}$ is as defined in Section~\ref{ccmod} and describes the central superconducting circuit. Here we also included applied voltages $V_{\text{g}k}$ (shown in Fig.~\ref{fig:schematic}a) that can lead to charge offset noise $N_{\text{g}k}$ on each island. 

\subsection{Single excitation S-matrix analysis}
\label{S3}

For this analysis, we work in the low-power, single excitation limit, wherein we consider only one photon as input to or output from the system and require that the excitation energy in the system only exist in one component of the system at any one time--- in one of the resonators, one of the transmission lines, or as an excited state of the central circuit. Upon enforcing this limit, we establish the S-matrix of the system using input-output theory. First, we write a generic state in this single excitation manifold as:
\begin{equation}
\label{eq:genstate}
\begin{split}
\ket{\Psi_0} \ = \ &\sum_{k=1}^3 d_k \hat{a}^{\dagger}_k \ket{G} \ + \ \sum_{n} c_n \ket{\varepsilon_n} \\
&+ \ \sum_{k=1}^3 \int d\nu f_k(\nu) \ \hat{b}^{\dagger}_k(\nu) \ket{G} \enspace ,
\end{split}
\end{equation}
where $d_k$, $c_n$, and $f_k(\nu)$ are time-dependent complex probability amplitudes for each state, and we abbreviate, with a slight abuse of notation, the ground state $\ket{0,0,0,G,0,0,0}$ as $\ket{G}$, and the excited circuit state with no photons in any resonator or transmission line $\ket{0,0,0,\varepsilon_n,0,0,0}$ as $\ket{\varepsilon_n}$. The summation over $n$ is a summation over only the excited states of the circuit--- it excludes the ground state as this state is not in the manifold being considered.

By using the Schr$\ddot{\text{o}}$dinger equation, we establish equations of motion for the time-varying complex probability amplitudes appearing in Eq.~(\ref{eq:genstate}). With these equations of motion, we can utilize input-output theory \cite{colgard, walls} to develop a relation connecting in-going and out-going mode amplitudes via the S-matrix. Proceeding with this analysis results in the usual input-output relation,
\begin{equation}
\label{eq:inoutS}
   \underline{f_{\text{out}}}(\omega) =\underline{\underline{S}}(\omega) \ \underline{f_{\text{in}}}(\omega) \enspace ,
\end{equation}
where $f_{\text{in}}(\omega)$ and $f_{\text{out}}(\omega)$ are the in-going and out-going single photon mode amplitudes in frequency space that reference either the initial or final time $t_\text{s}$ defined by:
\begin{equation}
\label{eq:INOUTdefs}
\frac{1}{\sqrt{2\pi}}\int d\nu \ e^{-i\nu(t-t_\text{s})}f_k(\nu,t_\text{s})=
\begin{cases}
    f_{\text{in},k}  & \text{if } t>t_\text{s}\\
    f_{\text{out},k} & \text{if } t<t_\text{s}
\end{cases} \enspace .
\end{equation}

The single photon S-matrix is found by solution in the frequency domain for the coefficients in our ansatz, $\ket{\Psi_0}$. In frequency space we have:
\begin{equation}
\label{eq:Smat}
\underline{\underline{S}}(\omega)=\mathbb{1}_{3\times3}+ \kappa \ \underline{\underline{\chi}}(\omega) \enspace ,
\end{equation}
where the inverse susceptibility of the resonators combined with the central circuit is given by:
\begin{equation}
\label{eq:nmat}
  \underline{\underline{\chi}}^{-1}(\omega)= [ \ i(\omega-\omega_\text{R})-\frac{\kappa}{2} \ ] \ \mathbb{1}_{3\times3}+g^2 \ \underline{\underline{T}}(\omega) \enspace .
\end{equation}
The elements of the term $\underline{\underline{T}}(\omega)$ are defined relative to starting in the ground state of the central circuit:
\begin{equation}
\label{eq:tmatex}
T_{jk}(\omega)= \sum_n \frac{1}{i \omega - i \varepsilon_n/\hbar} \bra{G}\hat{B}_j\ket{\varepsilon_n}\bra{\varepsilon_n}\hat{B}_k\ket{G} \enspace ,
\end{equation}
where we drop the tilde to indicate the complex-valued Fourier transform of the voltage-voltage linear response function given by Eq.~(\ref{eq:VV}) in Section~\ref{LR}. Having established the constituent pieces of this single photon S-matrix, we can now move forward to determine the necessary conditions for the desired ideal circulation.

\subsection{Conditions for ideal circulation}
\label{S4}

Having established the S-matrix using input-output theory, we would like to determine what conditions exist on parameters of the circuit such that this matrix matches that of an ideal circulator's S-matrix:

\begin{align}
\label{eq:ism}
        \underline{\underline{S_{\text{ccw}}}} &= \begin{bmatrix}
            0 & 1 & 0\\
    		0 & 0 & 1\\
			1 & 0 & 0
         \end{bmatrix}
&         
    \underline{\underline{S_{\text{cw}}}} &= \begin{bmatrix}
            0&0&1\\
    		1&0&0\\
			0&1&0
         \end{bmatrix} \enspace .
\end{align}

Examination of Eqs.~(\ref{eq:Smat}) and (\ref{eq:nmat}) indicates that any nonreciprocal behavior must originate from the circuit-specific term in the S-matrix, $\underline{\underline{T}}(\omega)$, as all other components are trivially diagonal. Focusing on this coupling matrix, we will now work to find conditions for nonreciprocity. Since the central circuit of our model is symmetric with respect to cyclic permutation, the diagonal elements of $\underline{\underline{T}}(\omega)$ are all equal, as are the off-diagonal elements that are cyclic permutations of each other, i.e., $T_{12}=T_{23}=T_{31}$ and $T_{21}=T_{32}=T_{13}$. With this in mind, we rewrite Eq.~(\ref{eq:tmatex}) as:
\begin{equation}
\label{eq:tmatc}
\underline{\underline{T}}(\omega) = -i \ \begin{bmatrix}
    \beta&\alpha&\alpha^*\\
    \alpha^*&\beta&\alpha\\
	\alpha&\alpha^*&\beta
    \end{bmatrix} \enspace ,
\end{equation}
with $\alpha = iT_{jj+1}$ and $\beta = iT_{jj}$.

We note that this problem can be solved exactly--- doing so reveals that optimal performance will occur when frequency shifts of the resonators due to the central circuit are compensated, and when the coupling to the central circuit is matched to the resonator decay. Explicitly, for a signal at target frequency $\omega_\text{T}$, these conditions are
\begin{equation}
\label{eq:detuningcon}
\omega_\text{R} = \omega_\text{T}-g^2 \beta \enspace ,
\end{equation}
and
\begin{equation}
\label{eq:kappacon}
\kappa = 2 |\alpha| g^2 \enspace .
\end{equation}

Rewriting $\underline{\underline{S}}(\omega_\text{T})$ for these optimal values, we have the following matrix equation:
\begin{equation}
    \underline{\underline{S}}(\omega_\text{T}) = \mathbb{1} - \frac{2}{\mathbb{1} + i \left( \begin{matrix}
    0 & e^{i \Delta \theta/6} & e^{-i \Delta \theta/6} \\
    e^{-i \Delta \theta/6} & 0 & e^{i \Delta \theta/6} \\
    e^{i \Delta \theta/6} & e^{-i \Delta \theta/6} & 0 \end{matrix} \right)} \enspace .
\end{equation}

This shows that near resonance ($\omega \approx \omega_\text{T}$), the S-matrix will be dominated by the behavior of $\underline{\underline{T}}(\omega)$. The first terms that contribute to nonreciprocal behavior can be understood by expanding $(\mathbb{1} + i  \eta \underline{\underline{T}}(\omega))^{-1}$ in $\eta < |T|$. We see terms in the series that correspond to clockwise or counterclockwise circulation beginning and ending at the same port, namely $T_{21}T_{32}T_{13}$ and $T_{12}T_{23}T_{31}$, respectively. If the central circuit is reciprocal, the quantity $|T_{12}T_{23}T_{31} \ - \ T_{21}T_{32}T_{13}|$, which represents the destructive interference between a signal going clockwise and counterclockwise around the circuit, should be zero. However, if the system is ideally nonreciprocal, they instead constructively interfere. Writing $\alpha = |\alpha| e^{i \Delta \theta/6}$, where the phase is written in terms of the difference accumulated between a clockwise and counterclockwise path $\Delta \theta$, we have $|T_{12}T_{23}T_{31} \ - \ T_{21}T_{32}T_{13}| = 2 |\alpha|^3 |\sin(\Delta \theta/2)|$. Thus for optimal behavior, 
\begin{equation}
\label{eq:phasediffcon}
\Delta \theta = 
\begin{cases}
     0, \pm2\pi... \enspace ,  & \text{ Reciprocal}\\
    \pm\pi, \pm3\pi... \enspace , & \text{ Nonreciprocal}
\end{cases} \enspace .
\end{equation}

We now have three conditions on our system that, when satisfied, yield ideal circulation in either the clockwise or counterclockwise direction:
\begin{equation}
\label{eq:NRcon}
\begin{split}
(\text{a})& \ \ \omega_\text{T} \ \text{s.t.} \ \Delta \theta = \pm\pi, \pm3\pi... \\ 
(\text{b})& \ \ \kappa = 2 |\alpha| g^2  \\
(\text{c})& \ \ \omega_\text{R} = \omega_\text{T}-g^2 \beta
\end{split} \enspace .
\end{equation}
With these conditions, as long as the central circuit has a target frequency for which the phase difference condition in Eq.~(\ref{eq:NRcon}a) holds, we may choose one of the external circuit parameters and let the conditions given in Eqs.~(\ref{eq:NRcon}b) or (\ref{eq:NRcon}c) determine the others. Note that these results are consistent with previous work, e.g., Ref. \cite{JOE}. Additionally, these results specify the phase differences corresponding to clockwise (CW) or counterclockwise (CCW) transmission, yielding:
\begin{equation}
\label{eq:PDdirection}
\Delta \theta =
\begin{cases}
     -\pi, 3\pi, -5\pi... \enspace , & \text{CW Transmission}\\
      \pi, -3\pi, 5\pi... \enspace, & \text{CCW Transmission}
\end{cases} \enspace .
\end{equation}
We now have all we need to explore our system.  

\section{Numerical results}
\label{S5}

We proceed now with a numerical analysis, examining the input-output properties of the entire system using the analytical results established in Sections~\ref{S3} and \ref{S4}. Broadly speaking, we wish to enforce the conditions for ideal circulation given in Eq.~(\ref{eq:NRcon}) and then examine the S-matrix given altogether by Eqs.~(\ref{eq:Smat}), (\ref{eq:nmat}), and (\ref{eq:tmatex}). As there are a variety of different regimes of the central circuit, we investigated several values of external flux $A$, described in Appendix~\ref{appending}. The best performing region was identified near $A \approx 0.2478$, and we focus on this in what follows.

To evaluate performance, we compute the elements of $\underline{\underline{T}}(\omega)$ [Eq.~(\ref{eq:tmatex})] using the energy eigenstates and eigenvalues established numerically in Section~\ref{VT2}. With the elements of $\underline{\underline{T}}(\omega)$, we find the phase difference $\Delta \theta$ as a function of potential incoming signal frequency $\omega$. Examining the phase difference allows us to identify target frequencies where $\Delta \theta = \pm\pi, \pm3\pi...$, and use Eqs.~(\ref{eq:NRcon}b) and (\ref{eq:NRcon}c) to set external circuit parameters that fix $\kappa,\ \omega_\text{R}$, and $g$. We can then examine relevant S-matrix properties, as well as introduce variation in the offset charge on each island and flux penetration per loop to test the system's susceptibility to charge and flux disorder, i.e., variations of these parameters on timescales well below the circuit's bandwidth.

\begin{figure}
\centering
\includegraphics[width = 0.48 \textwidth]{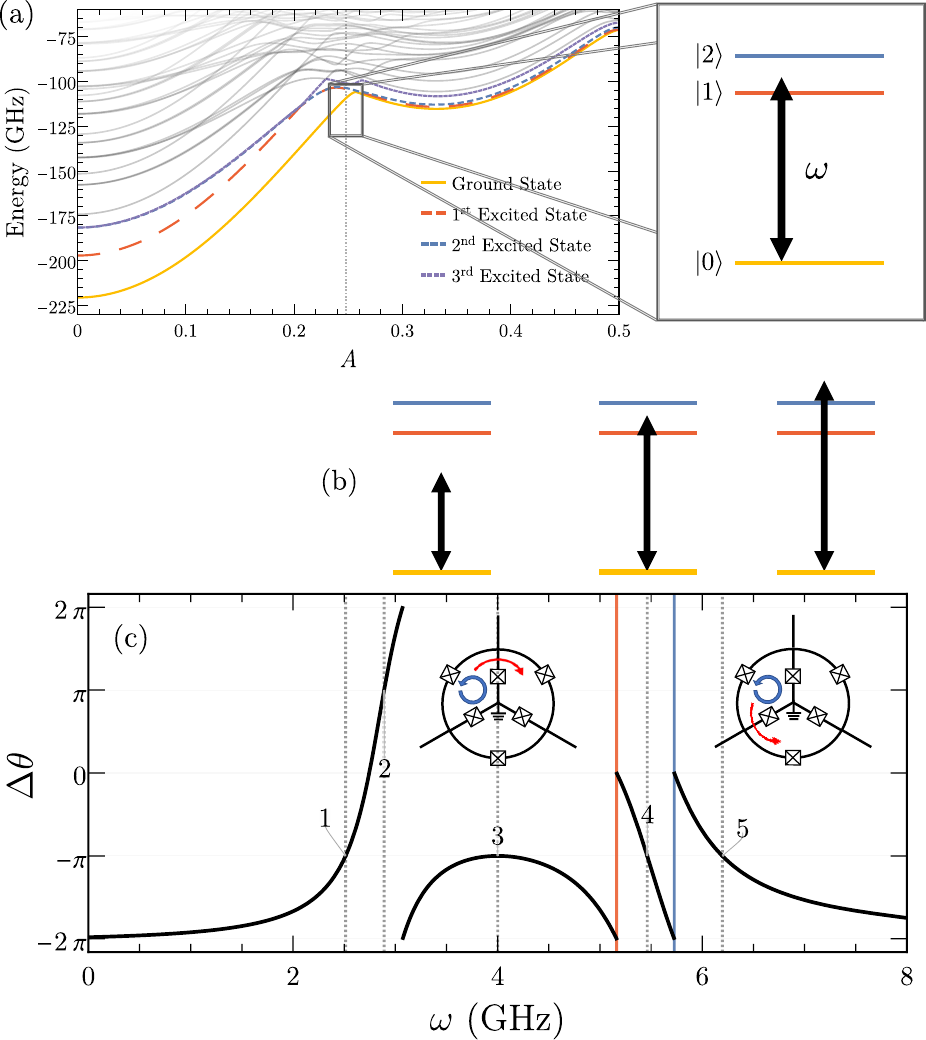}
\caption{In (a) and (b), we highlight the energy level structure at our chosen operating point $A \approx 0.2478$ and the available transitions from the ground state. The phase difference $\Delta \theta$ as a function of driving frequency $\omega$ in shown in (c), with insets that indicate the predominant vortex dynamics. Dashed vertical lines indicate the target frequencies at which $\Delta \theta = \pm \pi, \ \pm 3\pi...$--- the values that satisfy the  conditions for nonreciprocity. These points are enumerated for future reference. The red and blue vertical lines indicate the first and second excited energies, respectively,  for the central circuit. Parameters chosen for this plot are $E_\text{J}/h = 4 E_\text{C}/h \approx 45 \text{ GHz}$, where $E_\text{C} = \frac{e^2}{2C_\text{J}}$, $E_\text{J}/E_{\Sigma} \approx 40$, $C_\text{C} = 6 C_\text{J}$, $C_\text{G} = 0.02 C_\text{J}$, and $A \approx 0.2478$. Offset charge is tuned such that the total Cooper pair number in the ground state is $\approx 1.2$, with $N_{\text{g}1}=N_{\text{g}2}=N_{\text{g}3}\approx 0.4$.}
\label{fig:PDplot}
\end{figure}

There are numerous points for which the phase difference condition is satisfied, as indicated in Fig.~\ref{fig:PDplot}c. Of particular interest are instances like the third and fourth points, found at $\Delta \theta = -\pi$ where $\omega_\text{T3} \approx 4 \text{ GHz}$ and at $\Delta \theta = -\pi$ where $\omega_\text{T4} \approx 5.46 \text{ GHz}$, respectively. At the third point, the first derivative of the phase difference is zero, indicating much more stability with regard to frequency than at other target frequencies. At other points, the first derivative is much larger and a small change in frequency results in significantly larger changes to $\Delta \theta$. The fourth point is situated between the first and second excited energies of the central circuit. Due to the close proximity of this target frequency to two resonances of the circuit, it may be easier to inject energy into the circuit here than at other nonreciprocal points. In both cases, intuition suggests that these target frequencies, $\omega_\text{T3}$ and $\omega_\text{T4}$, may allow for increased bandwidth of the system. This possibility will be explored upon examination of elements of the S-matrix in Section~\ref{results}.

\subsection{Bandwidth, nonreciprocity, and circulation}
\label{results}

We now determine performance via the S-matrix property that the normalized output power at the $i^{\text{th}}$ port provided input at the $j^{\text{th}}$ port is given by $|S_{ij}|^2$. Setting the external circuit parameters $g$, $\kappa$, and $\omega_{\text{R}}$ for the center frequencies $\omega_{\text{T}i}$ with $i=1,2,3,4$, the target frequencies identified in Fig.~\ref{fig:PDplot}c, we fix $g$ to a reasonable but large value of $g=1.6$ GHz and use Eq.~(\ref{eq:NRcon}) to determine $\kappa$ and $\omega_{\text{R}}$. Table~\ref{tab:NRparameters} shows the resulting parameters for the first four target frequencies in Fig.~\ref{fig:PDplot}c. We remark that for large bandwidth operation, $g$ needs to be large, but making stronger coupling to the resonator has limits in other relevant parameters, as discussed in Appendix~\ref{app:PC}.

\setlength{\tabcolsep}{8pt}
\begin{table}
\caption{External Circuit Parameters for Nonreciprocal Points}
\vspace{1.0em}
\centering 
\begin{tabular}{c c c c c}
\hline\hline
Point & $\omega_{\text{T}}$ (GHz) & $g$ (GHz) & $\kappa$ (GHz) & $\omega_{\text{R}}$ (GHz) \\ [1ex] 
\hline 
1 & 2.5137 & 1.6 & 0.01321 & 2.7408\\ [1ex]
2 & 2.8904 & 1.6 & 0.00995 & 3.1343\\ [1ex]
3 & 4.0005 & 1.6 & 0.08998 & 4.3396\\ [1ex]
4 & 5.4618 & 1.6 & 1.67917 & 5.7508\\ [0.5ex]
\hline 
\end{tabular}
\label{tab:NRparameters}
\end{table}

\begin{figure*}
\centering
\includegraphics[width = 0.90 \textwidth]{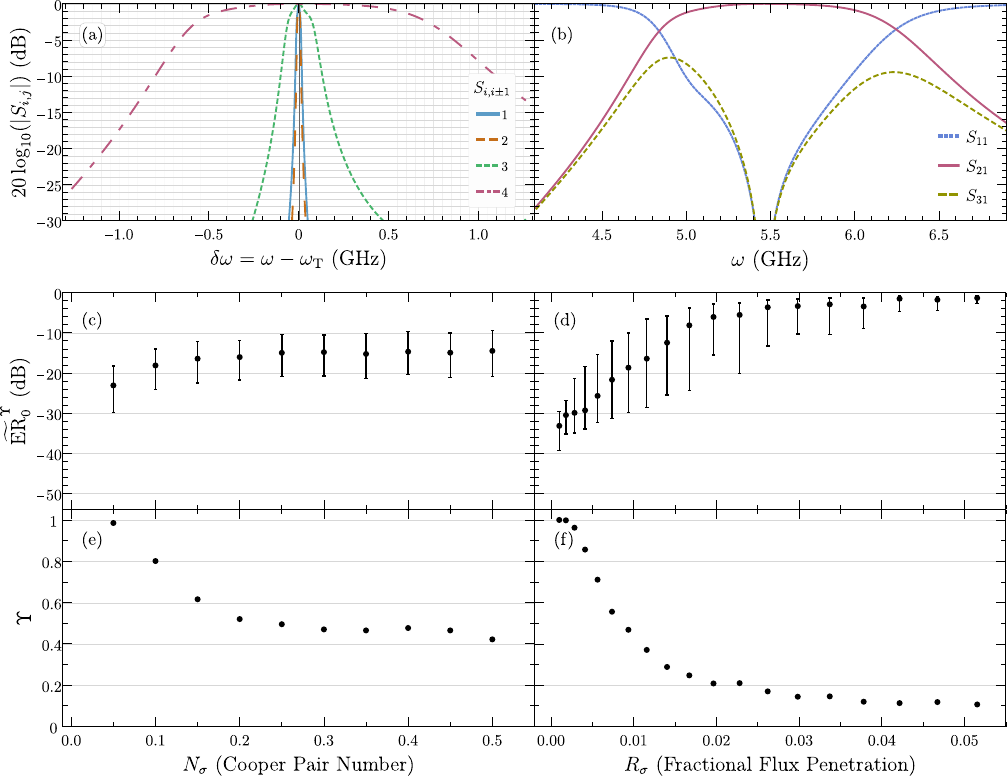}
\caption{(a) A comparison of the normalized output power at the $i^{\text{th}}$ port given input at the $i\pm1^{\text{th}}$ port for each set of external circuit parameters shown in Table \ref{tab:NRparameters}. For the first, second, and fourth points, transmission direction is such that $|S_{i,i-1}|^2$ is plotted. For the second point, $|S_{i,i+1}|^2$ is plotted. (b) A comparison of normalized output powers at each port assuming input at port one. At $\omega \approx 5.46 \text{ GHz}$ we observe full transmission in a clockwise fashion to the second port with zero transmission in the opposite direction, indicative of circulator behavior. Parameters and offset charge used are the same as in Fig.~\ref{fig:PDplot}. External circuit parameters are set according to the fourth point, shown in Table~\ref{tab:NRparameters}. The median of $\text{ER}^\Upsilon_\text{0}$ (defined in Appendix~\ref{app:SA} as the extinction ratio for devices with less than 1 dB of insertion loss), $\widetilde{\text{ER}}^\Upsilon_\text{0}$, is plotted as a function of the standard deviation of (c) the charge noise distribution $N_\sigma$ and (d) $R_\sigma$, the distribution of fractional flux penetration per loop. Error bars indicate the spread from the first to third quartile of $\text{ER}_\text{0}^\Upsilon$ for each value of $N_\sigma$ and $R_\sigma$. Device yield $\Upsilon$ (defined in Appendix~\ref{app:SA} as the fraction of devices with less than 1 dB of insertion loss) is plotted as a function of (e) $N_\sigma$ and (f) $R_\sigma$. Parameters for ideal tuning in plots (c)-(f) are the same as in~(b).}
\label{fig:num}
\end{figure*}

For each set of external circuit parameters specified by the target frequencies in Table~\ref{tab:NRparameters}, we wish to examine the bandwidth of the resulting transmission between two ports. For each case, the direction of maximum transmission is identified by the phase difference in Fig~\ref{fig:PDplot}c, and we plot the normalized output power (in dB), displaced by the target frequency of each for comparison. These results are shown in Fig.~\ref{fig:num}a.

We define the bandwidth as a 3 dB window: the width of the signal in frequency space over which transmission is greater than 50 \%. Examination of Fig.~\ref{fig:num}a indicates that the parameters for the first and second points result in bandwidths of approximately 23 MHz and 17 MHz, respectively. At the third and fourth points, however, we see much larger bandwidths. At the third point we find a bandwidth of approximately 144 MHz, while the parameter set for the fourth point yields a very large bandwidth of approximately 1.34 GHz. These considerably larger bandwidths appear to confirm the intuition described previously: features in the phase difference like that at the third point in Fig.~\ref{fig:PDplot}c can lead to increased bandwidth, but even larger bandwidth is achieved when nonreciprocal points are situated near resonances---e.g., the fourth point---providing greater ease in inserting energy into the system (that is, a large $|\alpha|$) and a larger range of frequencies over which it is possible to do so. Since this large bandwidth is desirable in systems such as these, we proceed now using the external circuit parameters for the fourth point (at $\Delta \theta = -\pi$ where $\omega_\text{T4} \approx 5.46 \text{ GHz}$).

Upon fixing external circuit parameters to maximize bandwidth, we examine nonreciprocal behavior and ideal circulation in the system. To do so, we compare transmission from a single port to each of the three outputs, plotting the normalized output powers (in dB) at each port provided input at a single port to observe any circulator behavior. Fig.~\ref{fig:num}b shows the resulting transmission to each port provided input at port one.

We see in Fig.~\ref{fig:num}b that microwave signals only propagate in one direction; there is a clear preferred direction of signal transmission, with full transmission at $\omega \approx 5.46 \text{ GHz}$ in the clockwise direction. Fig.~\ref{fig:num}b demonstrates the system is lossless and nonreciprocal, and at $\omega \approx 5.46 \text{ GHz}$, matched--- all characteristics of an ideal circulator's S-matrix \cite{pozar}. Note that due to the three-fold rotational symmetry of the system, these results are reproduced under all cyclic permutations.

\subsection{Resilience to charge and flux noise}
\label{SS:randomnoise}

Thus far, our analysis has assumed high symmetry: identical, static offset charges $N_{\text{g}i}$ on each island due to the presence of applied voltages $V_{\text{g}i}$ as well as identical loop areas in the central circuit (ensuring identical flux penetration per loop). In particular, given our expectation that charge offset is not a strong effect, we arbitrarily set $N_{\text{g}1}=N_{\text{g}2}=N_{\text{g}3} \approx 0.4$ Cooper pairs, $A \approx 0.2478$, and subsequently tuned the external circuit parameters for ideal circulation in the prior parts of this section. However, sources of noise that break this high symmetry are inevitable. Each island of the superconducting circuit may be subject to random variation in the offset charge that will necessarily affect performance given the intermediate energy regime in which we have chosen to work. Similarly, static flux variations across the device due to imperfections in manufacture of the central circuit or local impurities, unexpected magnetic field gradients, and other phenomena will likely degrade the ideal circulation we have observed in Section~\ref{results}. Therefore, we wish to address to what degree this is the case by introducing these sources of noise into our analysis. We note that an additional source of disorder is that of the Josephson junctions themselves; we examine this fabrication disorder in Appendix~\ref{app:FD}.

To test the system's susceptibility to local charge noise, we begin with the system tuned for ideal circulator behavior with equal, fixed offset charges $N_{\text{g}i}$. We then introduce an additional offset charge $\Delta_i$ that is unique for each island and randomly sampled from a normal distribution with a mean of $N_\mu = 0$ and a standard deviation $N_\sigma$ in units of Cooper pair number. As detailed and defined in Appendix~\ref{noiseapp}, we assess performance in the presence of charge noise by examining $\text{ER}^\Upsilon_\text{0}$ and $\Upsilon$ as a function of $N_\sigma$, shown in Figs.~\ref{fig:num}c and \ref{fig:num}e, respectively. These quantities convey the distribution of the minimum of the extinction ratio for those noise samples resulting in less than 1 dB of insertion loss and the likelihood that this is the case using a sample size of 1200 noise samples.

Upon increasing $N_\sigma$, we observe an initial slight decrease in performance indicated by the increase in $\widetilde{\text{ER}}^\Upsilon_\text{0}$ (the median of the distribution of $\text{ER}^\Upsilon_\text{0}$), as shown in Fig.~\ref{fig:num}c. However, at $N_\sigma \approx 0.25$ Cooper pairs we observe saturation to $\widetilde{\text{ER}}^\Upsilon_\text{0} \approx$ -15 dB. This saturation is reasonable given that the behavior of the circuit is unchanged upon rescaling by any integer number of Cooper pairs. Once $N_\sigma$ exceeds $\approx 0.25$, the distribution is wide enough to encompass this rescaling. The device yield $\Upsilon$ displays similar behavior with increasing $N_\sigma$, as shown in Fig.~\ref{fig:num}e, where the likelihood of large transmission also saturates at $\approx 0.25$ to about 50 \%. It is very promising that even in the presence of essentially arbitrary, static charge noise, the system retains its nonreciprocity.

To test the system's susceptibility to flux noise, we begin with the system tuned for ideal circulator behavior where all three loops of the central circuit are equal in area and penetrated by a flux $A \approx 0.2478$. We then introduce variation in the flux penetration of each loop, $\Delta_i'$, that is unique for each loop but maintains the total area and flux penetration of the ideal case, i.e., $\Delta_1'+\Delta_2'+\Delta_3'=0$, to account for tuning the flux optimally for a given device (where the total field can be varied but not, perhaps, the individual loop fields). We take $\Delta_i'$ to be a fraction of $A$, the original flux penetration per loop, randomly sampling from a normal distribution with a mean of $R_\mu = 0$ and a standard deviation $R_\sigma$. As detailed and defined in Appendix~\ref{noiseapp}, we assess performance in the presence of flux noise by examining $\text{ER}^\Upsilon_\text{0}$ and $\Upsilon$ as a function of $R_\sigma$, shown in Figs.~\ref{fig:num}d and \ref{fig:num}f, respectively. As stated above, $\text{ER}^\Upsilon_\text{0}$ and $\Upsilon$ convey the distribution of the minimum of the extinction ratio for those noise samples resulting in less than 1 dB of insertion loss and the likelihood that this is the case using a sample size of 1200 noise samples.

Upon increasing $R_\sigma$, we observe a decrease in performance indicated by the increase in $\widetilde{\text{ER}}^\Upsilon_\text{0}$ (the median of the distribution of $\text{ER}^\Upsilon_\text{0}$), as shown in Fig.~\ref{fig:num}d. This degradation in performance is more substantial than in the charge noise case--- decreasing more rapidly and saturating at $R_\sigma \approx 0.025$ to $\widetilde{\text{ER}}^\Upsilon_\text{0} \approx$ -1 dB. Note that this flux noise model encompasses both random variations in loop area, e.g., imperfections in manufacture, as well as static flux variations across the device that may occur. We do not consider variation in total flux penetration, as this quantity can be experimentally calibrated.

\section{Conclusions}
\label{conclusion}

In this work, we have presented a theoretical analysis of a compact superconducting circuit operating as a circulator. Our investigation shows that operating in a non-charge-conserved regime with mobile vortices in the XY model, where the charging energy and Josephson energy scales are similar, leads to performance with a reduced dependence on static charge noise. Furthermore, the presence of both clockwise and counterclockwise vortex circulation at nearby energies appears to enable wide bandwidth performance that exceeds one GHz. 

However, a variety of open questions remain for further investigation. It is likely possible to generalize the understanding achieved in this three-port model to a circulator with greater than three ports, e.g., a double-junction circulator, which could be explored in future work. In addition, in the context of quantum computing and hardware, circulators are typically needed to handle both lower power signal routing as well as higher power device isolation. In this work, our single photon analysis suggests the model is suitable for such low power signal routing. However, a crucial operational question is the performance of the system at higher powers, beyond the single photon limit, thereby expanding the model's potential use cases. We anticipate that further exploration of the vortex-tunneling picture may elucidate regimes of operation where photon-vortex-photon terms enable high power operation, but may also require more complex circuit designs. Another key question is whether ordered vortex states, such as those arising in larger arrays, can exhibit robust edge modes for vortex circulation that may further reduce the effects of noise and disorder in circulator performance. Additionally, implementation of this type of system in an experimental setting will help improve our understanding of quantum vortex dynamics and may lead to other new avenues of exploration (e.g., Ref.~\cite{weyl}) for these and other topological excitations.

\begin{acknowledgments}
We thank Y. Nakamura, C. Raj, H. Ikegami, C. Lobb, V. Manuchuryan, and V. Galitski for helpful conversations. This work was supported by the NSF-funded Physics Frontier Center at the Joint Quantum Institute and by the Princeton Center for Complex Materials NSF-funded MRSEC.
\end{acknowledgments}

\bibliography{REFS}

\newpage
\appendix

\section{Alternative circuit--- floating central island}
\label{app:float}

The implementation of the grounded node present in the proposed central circuit (as shown in Fig.~\ref{fig:schematic}b) is a difficult experimental task. A simple modification---floating the central node---provides an alternative circuit that is easier to realize. In this appendix, we explore this alternative circuit (as shown in Fig.~\ref{fig:schematic}c), highlighting the connection between this model and the original and showing that similar behavior can be achieved without the presence of a central grounded node.

By replacing the central ground with a floating island, we introduce two additional degrees of freedom--- the charge number and phase on this new island, $\hat{N}_0$ and $\hat{\varphi}_0$, respectively. To achieve similarity with the original model, the new central island should mimic a ground node, which we can accomplish by increasing the island's capacitance via its physical size. However, in doing so, we can no longer neglect the self-capacitance of the now large central island as well as the associated mutual capacitance it has with the outer three islands. 

In all, the modified Hamiltonian is given by:
\begin{equation}
\begin{split}
H = &\frac{(2e)^2}{2} \ (\underline{\hat{N}} - \underline{N}_\text{g})^\text{T} \ \underline{\underline{C}}^{-1} \ (\underline{\hat{N}} - \underline{N}_\text{g}) \\
 &- E_\text{J} \sum_{i=1}^3 \left[ \cos(\hat{\varphi}_i-\hat{\varphi}_0) + \cos(\hat{\varphi}_{i+1}-\hat{\varphi}_i-2 \pi A) \right] 
\end{split} \enspace ,
\label{eq:newH}
\end{equation}
where now $\underline{\hat{N}} = \lbrace \hat{N}_1,\hat{N}_2, \hat{N}_3,\hat{N}_0 \rbrace$ and $\underline{N}_\text{g} = \lbrace N_\text{g1},N_\text{g2},N_\text{g3},0 \rbrace$. The capacitance matrix $\underline{\underline{C}}$ is now a 4x4 matrix with the form:
\begin{equation}
\underline{\underline{C}} = \begin{bmatrix}
    C'_\text{S}&-C_\text{J}&-C_\text{J}&-C_\text{F}\\
    -C_\text{J}&C'_\text{S}&-C_\text{J}&-C_\text{F}\\
    -C_\text{J}&-C_\text{J}&C'_\text{S}&-C_\text{F}\\
    -C_\text{F}&-C_\text{F}&-C_\text{F}&C_0+3C_\text{F}
    \end{bmatrix} \enspace ,
\label{eq:newCmat}
\end{equation}
where $C'_\text{S} = C_\text{C}+C_\text{G}+C_\mu+3C_\text{J}$ and $C_\text{F} = C_\text{J}+C_\mu$ with $C_0$ the self-capacitance of the center island, $C_\mu$ the mutual capacitance between it and the outer islands, and all other capacitances as defined in Section~\ref{ccmod} of the main text. We express the eigenvalues of the capacitance matrix in Eq.(\ref{eq:newCmat}) as
\begin{equation}
\begin{split}
&\lambda_1=\lambda_2 = C_\text{C}+C_\text{G}+C_\mu+4C_\text{J} \\
&\begin{split}\lambda_3\lambda_4 = C_0(C_\text{C}&+C_\text{G}+C_\mu+C_\text{J})\\
&+3(C_\mu+C_\text{J})(C_\text{C}+C_\text{G}) \end{split}
\end{split} \enspace ,
\label{eq:newEV}
\end{equation}
and define the quantity $\lambda'=C_0 C_\text{J} + (C_\text{J}+C_\mu)(4C_\text{J}+C_\mu)$ so that we may write the inverse capacitance matrix $\underline{\underline{C}}^{-1}$ in the compact form:
\begin{equation}
\underline{\underline{C}}^{-1} = \frac{1}{\lambda_1} \begin{bmatrix}
    1+\delta'&\delta'&\delta'&\eta\\
    \delta'&1+\delta'&\delta'&\eta\\
    \delta'&\delta'&1+\delta'&\eta\\
    \eta&\eta&\eta&\xi
    \end{bmatrix} \enspace ,
\label{eq:newICmat}
\end{equation}
where $\delta'= \frac{\lambda'}{\lambda_3 \lambda_4}$, $\eta=\frac{\lambda_1 (C_\text{J} + C_\mu)}{\lambda_3 \lambda_4}$, and $\xi=\frac{\lambda_1 (\lambda_1-3C_\text{J})}{\lambda_3 \lambda_4}$. 

Notably, by replacing the central ground with a floating island, this alternative central circuit is now charge-conserving. We therefore define the total charge operator $\hat{N}_\text{T} = \hat{N}_0 + \sum_{i=1}^3 \hat{N}_i$, which commutes with the Hamiltonian $[H,\hat{N}_\text{T}]=0$ and whose expectation value is constant, yielding the total charge in the system $\langle \hat{N}_\text{T} \rangle = n$. With this conserved quantity, we can eliminate a dynamical degree of freedom. 

In particular, by considering the unitary transformation
\begin{equation}
\hat{U} = e^{-i \sum_{i=1}^3 \hat{N}_i \hat{\varphi}_0} \enspace ,
\label{eq:UT}
\end{equation}
we can eliminate the charge number and phase operators of the central island from the Hamiltonian. Under this transformation, the charge number operators of the outer islands are unaffected while the charge number operator of the central island transforms as $\hat{N}_0 \rightarrow \hat{N}_0 - \sum_{i=1}^3 \hat{N}_i$. Similarly, the phase operators of the outer islands transform as $\hat{\varphi}_i \rightarrow \hat{\varphi}_i  + \hat{\varphi}_0$ while that of the central island is unchanged. Note that the total charge operator transforms as $\hat{N}_\text{T} \rightarrow \hat{N}_0$, i.e., in this new frame $\hat{N}_0$ represents total charge and is the conserved quantity. 

Performing the unitary transformation in Eq.(\ref{eq:UT}) on the Hamiltonian in Eq.(\ref{eq:newH}), we find that the Hamiltonian in this frame takes a form 
\begin{equation}
\begin{split}
H' = 4 &E'_\Sigma \ (\underline{\hat{N}} - \underline{N}'_\text{g})^\text{T} \ \underline{\underline{K'}}^{-1} \ (\underline{\hat{N}} - \underline{N}'_\text{g})\\
&- E_\text{J} \sum_{i=1}^3 \left[ \cos(\hat{\varphi}_i) + \cos(\hat{\varphi}_{i+1}-\hat{\varphi}_i-2 \pi A) \right] 
\end{split} \enspace ,
\label{eq:newHNF}
\end{equation}
where $\underline{\hat{N}} = \lbrace \hat{N}_1,\hat{N}_2,\hat{N}_3 \rbrace$, $\underline{N}'_\text{g} = \lbrace N'_\text{g1},N'_\text{g2},N'_\text{g3} \rbrace$. The form of the Hamiltonian in this frame is identical to that of our original model [$H_\text{CC}$ given by Eq.(\ref{eq:cc})], where now the charging energy of the circuit $E'_\Sigma$, its dimensionless inverse capacitance matrix $\underline{\underline{K'}}^{-1}$, and the offset charges $N'_{\text{g}i}$ all take on slightly modified forms. $E'_\Sigma$ is defined as $E'_\Sigma=\frac{e^2}{2 \lambda_1}$, where $\lambda_1= C_\text{C}+C_\text{G}+C_\mu+4C_\text{J}$ plays the role of $C_\text{M}$ from the main text but modified by the presence of $C_\mu$. We find that  $\underline{\underline{K'}}^{-1}$ takes a familiar form\begin{equation}
\underline{\underline{K'}}^{-1} = \begin{bmatrix}
    1+\rho&\rho&\rho\\
    \rho&1+\rho&\rho\\
    \rho&\rho&1+\rho
    \end{bmatrix} \enspace ,
\label{eq:newK}
\end{equation}
where $\rho = \delta'-2\eta+\xi$ and we see upon comparison to Eq.(\ref{eq:kimat}), we just have $\delta \rightarrow \rho$. Finally, the new offset charges are uniformly shifted from their original value:
\begin{equation}
N'_{\text{g}i} = N_{\text{g}i} - \frac{\eta-\xi}{1+3\rho} \left(\hat{N}_0-\sum_{j=1}^3 N_{\text{g}j}\right) \enspace .
\label{eq:newNG}
\end{equation}
Note that in the limit where $C_0 \rightarrow \infty$ and $C_\mu \rightarrow 0$, these modified components revert to the forms specific to $H_\text{CC}$ (the original model's Hamiltonian). However, provided a relatively large $C_0$ and small $C_\mu$ (e.g., $C_0 = 30 C_\text{J}$ and $C_\mu = 0.25 C_\text{J}$ suffice), we find similar transmission performance in a zero-noise setting as compared with that of the original circuit with a central ground. Importantly, we see that aside from the more straightforward parameter-related modifications to $E'_\Sigma$ and $\underline{\underline{K'}}^{-1}$, the presence of this new floating central island is equivalent to our original model with the addition of an offset charge that acts symmetrically on each island.

\begin{figure}
\centering
\includegraphics[width=0.48 \textwidth]{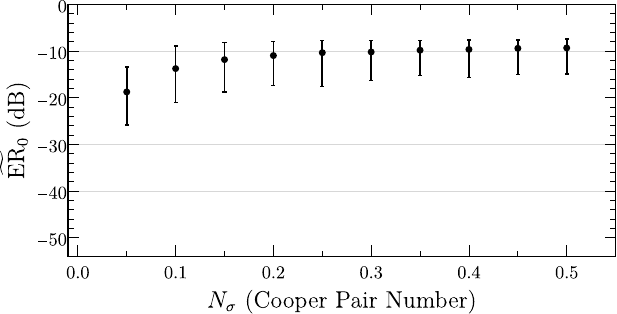}
\caption{Uniform charge disorder in the original model. The median of $\text{ER}_\text{0}$, $\widetilde{\text{ER}}_\text{0}$, is plotted as a function of the standard deviation of the charge noise distribution $N_\sigma$. Error bars indicate the spread from the first to third quartile of $\text{ER}_\text{0}$ for each value of $N_\sigma$. Parameters for ideal tuning are the same as in Fig.~\ref{fig:num}b.}
\label{fig:uniformnoise}
\end{figure}

In Section~\ref{SS:randomnoise} of the main text, we examined the effects of random charge disorder on performance by introducing a shift in offset charge that was randomly sampled and unique to each island. Having uncovered the equivalence between our original model and this alternative circuit, we can examine the effect of replacing the central ground with a floating island by instead looking at uniform charge disorder, i.e., introducing a shift in offset charge that is randomly sampled but identical for each island, in the original circuit with a central ground. This is shown in Fig.~\ref{fig:uniformnoise} without implementing any transmission threshold (as described Appendix~\ref{app:SA}). 

Regarding performance, we see in Fig.~\ref{fig:uniformnoise} a saturation to $\widetilde{\text{ER}}_\text{0} \approx -10$ dB at $N_\sigma \approx 0.25$. We can understand the increase in $N_\sigma$ displayed on the x-axis in Fig.~\ref{fig:uniformnoise} as an increase in total charge $N_0$, since the uniform shift we introduce is linearly dependent on $N_0$ as shown by Eq.(\ref{eq:newNG}). Generally, this total charge can be taken to be a constant scalar quantity, as it is typically fixed to some value in the process of cooling down the system in an experimental setting. However, these results suggest continued performance even with the addition of a central island and some variation in $N_0$. While it may be possible in an experimental setting to compensate for the shift in offset charge due to the additional island, it is promising that without any intervention, we find good performance.

\section{Investigation of additional external flux regions}
\label{appending}

\begin{figure}
\centering
\includegraphics[width = 0.48 \textwidth]{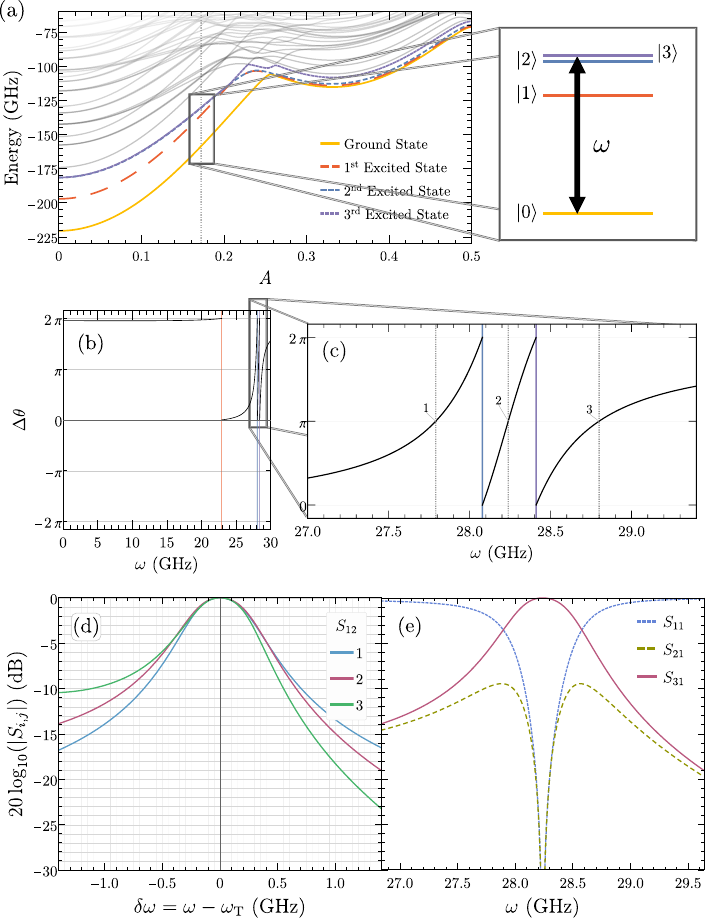}
\caption{In (a), we highlight the energy level structure at $A=0.1725$ in region 1 and the relevant transition from the ground state. The phase difference $\Delta \theta$ as a function of driving frequency $\omega$ is shown below in (b) and (c), where dashed vertical lines indicate the nonreciprocal target frequencies, enumerated for future reference. The red, blue, and purple vertical lines indicate the first, second, and third excited energies, respectively, for the central circuit. Parameters for these plots are as in Fig.~\ref{fig:PDplot} of the main text, except with $A=0.1725$. For the enumerated nonreciprocal points in (c), we show (d) transmission bandwidth and (e) signal circulation. Parameters are as in (b) and (c), where the procedure for determining $g$, $\kappa$, and $\omega_{\text{R}}$ is identical to that described Section~\ref{results} in the main text.}
\label{fig:A1}
\end{figure}

\begin{figure}[H]
\centering
\includegraphics[width = 0.48 \textwidth]{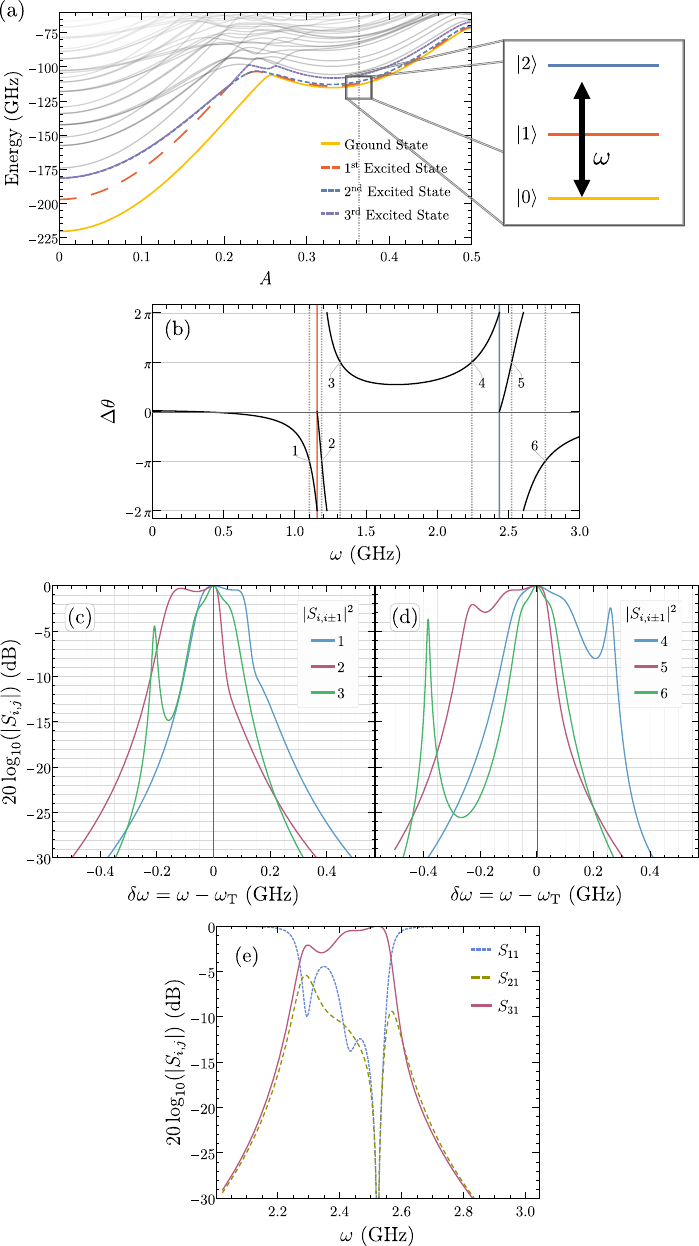}
\caption{In (a), we highlight the energy level structure at $A \approx 0.36$ in region 3. The phase difference $\Delta \theta$ as a function of driving frequency $\omega$ is shown below in (b), where dashed vertical lines indicate the nonreciprocal target frequencies, enumerated for future reference. The red and blue vertical lines indicate the first and second excited energies, respectively, for the central circuit. Parameters for these plots are as in Fig.~\ref{fig:PDplot} of the main text, except with $A \approx 0.36$. For the enumerated nonreciprocal points in (b), we show transmission bandwidth in (c) and (d). For point 5, we show signal circulation in (e). Parameters are as in (b), where the procedure for determining $g$, $\kappa$, and $\omega_{\text{R}}$ is identical to that described Section~\ref{results} in the main text.}
\label{fig:A2}
\end{figure}

Upon examination of the energy levels of the central circuit as a function of $A$, shown in Fig.~\ref{fig:ELamps}a of the main text, we identify three qualitative regions of behavior occurring from $ 0 \leq A \lesssim 0.2$ (region 1), $0.2 \lesssim A \lesssim 0.26$ (region 2), and $0.26 \lesssim A \leq 0.5$ (region 3). In the main text, we present the results of an analysis from region 2, where $A \approx 0.2478$. Here we investigate two additional flux values, $A \approx 0.17 \text{ and } A \approx 0.36$, in the first and third regions, respectively (highlighted in Figs.~\ref{fig:A1}a and \ref{fig:A2}a). In contrast to the large bandwidth circulator performance found at $A \approx 0.2478$, we find that the chosen points in the first and third regions do not exhibit similar qualities. We understand this behavior as follows.

In region 1, we note that the second and third excited states are nearly degenerate while the first excited state maintains a nearly constant separation from the ground state, mirroring the ground state, as evidenced in Figs.~\ref{fig:ELamps}a and \ref{fig:ELamps}b of the main text. Furthermore, in this region there are no circulating vortices between the ground and first excited states, in contrast to that with the second excited state, shown in Figs.~\ref{fig:directions}d and \ref{fig:directions}e. Since the expected mode of circulator operation is via the tunneling of persistent current vortices, points for ideal transmission should not occur around the first excited state. This is confirmed by examining the phase difference at $A \approx 0.17$, shown in Figs.~\ref{fig:A1}b and \ref{fig:A1}c, where ideal nonreciprocal points do not appear until the second and third excited states. This is understood by noting that between the ground and both the second and third excited states there exists circulating and counter-circulating vortices, analogous to the vortex dynamics of the first and second excited states in region 2. We also note that at $A \approx 0.17$, the energy level structure of the first, second, and third excited states (highlighted in Fig.~\ref{fig:A1}a) is similar to that of the ground, first excited, and second excited states at $A \approx 0.2478$ in region 2 (as investigated in the main text). However, the separation of the second and third excited states at $A \approx 0.17$ is about half the size of the separation between the first and second excited states at $A \approx 0.2478$. Examining performance at $A \approx 0.17$, we find transmission is relatively large around the second and third excited states, shown in Figs.~\ref{fig:A1}d and \ref{fig:A1}e, as compared to transmission at higher energy nonreciprocal points (not shown here). However, due to the closer proximity of the second and third excited states (as compared to the first and second excited states at $A \approx 0.2478$), we find smaller bandwidth when compared to that found at $A \approx 0.2478$. 

In region 3, the energy level structure is markedly different than either of the other two regions. As shown in Fig.~\ref{fig:A2}a, the level separations between the ground state, first excited state, and second excited state are roughly equivalent, and higher energy states are overall more accessible. Vortices are present between the ground and both the first and second excited states (as well as others). In fact, the first and second excited states exhibit circulating and counter-circulating vortices that are qualitatively similar to that of the second region. For this reason, numerous points for ideal circulation around these resonances should occur, which is confirmed by the phase difference found at $A \approx 0.36$, shown in Fig.~\ref{fig:A2}b. However, despite the abundance of vortex dynamics and overall lower energy as compared to either the first or second regions, the resulting performance---shown in Figs.~\ref{fig:A2}c, \ref{fig:A2}d, and \ref{fig:A2}e---is not as good as that found in region 2. This is perhaps due to the larger separation of energy levels, in particular between the first and second excited states, in contrast to the first and second excited states in region~2.

\section{Parameter Considerations}
\label{app:PC}

In the model we have considered, shown in Figs.~\ref{fig:schematic}a and \ref{fig:schematic}b, there are a variety of central circuit and external circuit parameters that can affect system behavior. Unfortunately, not all of these parameters have clear analytic expressions that characterize their influence on device performance. When considering performance metrics like transmission bandwidth and noise sensitivity, we must strike a balance between these parameters to find optimum performance. In this appendix, we discuss some of these parameters and our understanding of their effect on the system's performance.

Regarding the central superconducting circuit, of tremendous influence is the ratio of Josephson energy $E_\text{J}$ to charging energies $E_\text{C}$ and $E_\Sigma$. As touched on in Section~\ref{VT1}, the relative size of these energies impacts the relevant excitations and their quantum behavior, however, it also impacts the circuit's noise sensitivity. As $E_\text{J}$ increases with fixed plasma frequency ($E_{\text{C}},E_{\Sigma}$ decreasing together), the circuit becomes more robust against charge noise, but more susceptible to flux noise. The opposite is true as $E_\text{J}$ decreases. Additionally, our investigation suggests that this ratio impacts the role of fabrication disorder, as discussed in Appendix~\ref{app:FD}, whereby increasing $E_\text{J}$ leads to more susceptibility to noise in $E_\text{J}$ and less susceptibility to noise in various capacitances. Another important feature is the effect $E_\text{J}$ has on the separation of adjacent energy levels of the central circuit. As indicated by our investigations in Appendix~\ref{appending}, we believe that the proximity of relevant energy levels impacts the bandwidth achieved in each region of flux. As $E_\text{J}$ increases with fixed plasma frequency, the separation of the circuit's energy levels decreases, limiting bandwidth performance. Considering all of these effects together, our goal with the results presented in the main text, where $E_\text{J}/E_{\text{C}}=4$ and $E_\text{J}/E_{\Sigma} \approx 40$, is to balance the benefits of increased bandwidth with the detriment of decreased resistance to charge noise, all while remaining in a regime where the quantum behavior of vortices may be exploited. 

In addition to central circuit parameters, the parameters that characterize the connected external circuits ($g$, $\kappa$, $\omega_\text{R}$) are also quite important, influencing bandwidth performance as well as placing physical constraints on the elements of the system. Generally speaking, increasing the size of either of the coupling strengths $g$ or $\kappa$ will provide some increase in bandwidth. However, doing so affects physically relevant quantities such as the capacitance and characteristic impedance of the resonators ($C_\text{R}$,$Z_\text{R}$) and transmission lines ($C_\text{TL}$,$Z_\text{TL}$). A microscopic derivation utilizing the weak coupling limit ($C_\text{C} \ll C_\text{R},C_\text{TL}$), as done in Refs.\cite{koch,MDstace}, yields expressions for both $g$ and $\kappa$, the coupling strength between the circuit and resonators and the coupling strength between the resonators and transmission lines, respectively. For $g$ we find
\begin{equation}
\label{eq:g}
g =  \omega_\text{R} \frac{C_\text{C}}{C_\text{M}} \sqrt{\frac{8 Z_\text{R}}{R_\text{K}}} \enspace ,
\end{equation}
where $\omega_\text{R}$ is the resonator frequency, $Z_\text{R}$ the characteristic impedance of the resonators, and $R_\text{K}$ the von Klitzing constant. The capacitances $C_\text{C}$ and $C_\text{M}$ are as defined in Section~\ref{ccmod} of the main text. Increasing the size of $g$ (and necessarily $\omega_\text{R}$) increases $Z_\text{R}$ while pushing the ratio $C_\text{R}/C_\text{C}$ down, impacting the validity of the weak coupling limit. When combined with the conditions for ideal circulation given by Eq.~(\ref{eq:NRcon}) and the resulting parameter specifications given in Table~\ref{tab:NRparameters}, Eq.~(\ref{eq:g}) determines the characteristic impedance of the resonators required to achieve ideal circulation for each operating point. In the results discussed in Section~\ref{results} and shown in Fig.~\ref{fig:num}, the coupling strength was chosen to be $g=1.6$ GHz for all operating points. For the fourth operating point, which has particularly large bandwidth, this value of $g$ along with the other parameters given in the main text can be used in Eq.~(\ref{eq:g}) to give a value of $Z_\text{R} \approx 700 \ \Omega$--- larger than that of a typical coplanar waveguide, but not unreasonable \cite{HIres1,HIres2,HIres3}. 

In addition, we find for $\kappa$ the expression
\begin{equation}
\label{eq:kappa}
\kappa = 2 \pi \omega_\text{T}  \left( \frac{C_\text{C}}{C_\text{R}} \right)^2 \frac{Z_\text{TL}}{Z_\text{R}} \enspace ,
\end{equation}
where $\omega_\text{T}$ is the target frequency, $Z_\text{TL}$ the characteristic impedance of the transmission lines, and $C_\text{R}$ the capacitance of the resonators. Increasing $\kappa$, given the required conditions for ideal circulation as provided in Eq.~(\ref{eq:NRcon}) of the main text, necessarily increases $g$, resulting in the same impact on $Z_\text{R}$ and $C_\text{R}/C_\text{C}$ as described above. Therefore, while gains in bandwidth may be possible by increasing $g$ or $\kappa$, it must be weighed against the resulting increases in these physically constraining parameters. 

We can also use Eq.~(\ref{eq:kappa}) and the parameter values shown in Table~\ref{tab:NRparameters} of the main text to obtain the required transmission line impedance for an increased $g$. Unsurprisingly, given the size of $\kappa$ needed for a larger $g$, we also need the line impedance tuned to enable this strong coupling, resulting in $Z_\text{TL} \approx 5 \ \text{k}\Omega$. While this is very large, this is a microwave engineering challenge that may have various solutions, such as inductively coupling, rather than capacitively coupling, the transmission lines to the resonators. Despite this, given the numerous parameters (and the necessarily vast parameter space), there may exist as-yet-undiscovered operating points that maintain or improve upon our reported bandwidth and noise performance but decrease these large impedance requirements. Our investigation indicates that the presence of nonreciprocal behavior relies on the tuning of parameters relevant to the central superconducting circuit. While the external circuit parameters place physical constraints on the system, we still find that nonreciprocity can be achieved with any reasonable set of $g$, $\kappa$, and $\omega_\text{R}$. We note that while in this work, we opted to first fix $g$ and then examine requirements of the resonators, one could alternatively first choose the properties of the resonators and examine the resulting implications on $g$ and other system parameters.

\section{Details of the Noise Investigations}
\label{noiseapp}

We begin by defining an extinction ratio ER that compares the transmission in each direction between two ports of the circuit:
\begin{equation}
\label{eq:ER}
\text{ER}= 
\begin{cases}
20 \ \log_{10}\left( |\frac{S_{ij}}{S_{ji}}|\right) & \text{if } |S_{ij}| < |S_{ji}| \\
20 \ \log_{10}\left( |\frac{S_{ji}}{S_{ij}}|\right) & \text{if } |S_{ij}| > |S_{ji}|
\end{cases} \enspace .
\end{equation}
This metric provides a comparison between the desired direction of transmission and the undesired, reverse direction of transmission. For an ideal circulator, $\text{ER}\rightarrow -\infty$, thus in practice, the more negative this quantity, the more nonreciprocal the device. The minimum of the extinction ratio $\text{ER}_{\text{0}}$ (as a function of frequency) is what we will use to assess the system's response to noise.

Similarly, we define the insertion loss IL between two ports of the circuit as
\begin{equation}
\label{eq:IL}
\text{IL}= 
\begin{cases}
20 \ \log_{10}\left( |S_{ij}| \right) & \text{if } |S_{ij}| > |S_{ji}| \\
20 \ \log_{10}\left( |S_{ji}| \right) & \text{if } |S_{ij}| < |S_{ji}|
\end{cases} \enspace .
\end{equation}
The insertion loss quantifies the amount of transmission lost between two ports of the circuit. For zero transmission, $\text{IL}\rightarrow -\infty$. However, in an ideal circulator there is full transmission between two ports of the system, resulting in $\text{IL} = 0$. To help characterize the system's response to noise, we will examine the maximum of the insertion loss $\text{IL}_{\text{0}}$, where the closer this value is to zero, the better the performance.

\begin{figure}
\centering
\includegraphics[width = 0.48 \textwidth]{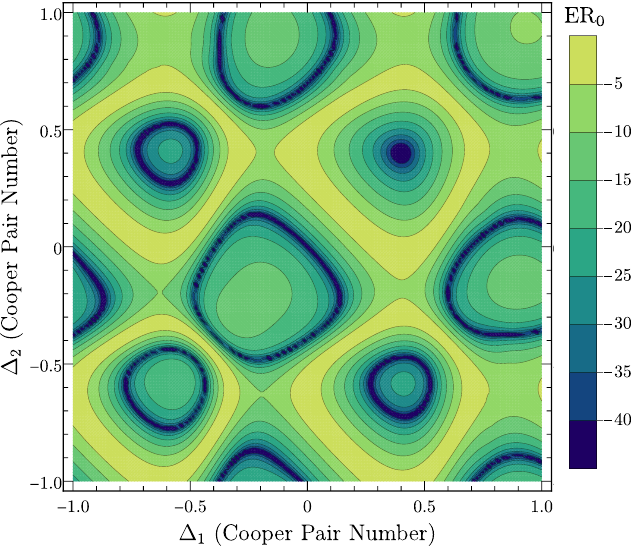}
\caption{The effect of offset charge number variations from the ideal on the minimum of the extinction ratio $\text{ER}_{\text{0}}$ (in dB) between ports one and two. Shown is the two-dimensional slice for $\Delta_3 = 0$, that is, an unchanged $N_{\text{g}3}$, while introducing variations $\Delta_1$ and $\Delta_2$ from $-1$ to $+1$ in $N_{\text{g}1}$ and $N_{\text{g}2}$, respectively. Parameters and offset charge for ideal tuning are the same as in Fig.~\ref{fig:PDplot}.}
\label{fig:A6}
\end{figure}

The introduction of deviations from ideal parameters (noise) will necessarily lead to variations in device performance. These deviations parameterize a multidimensional landscape of device performance, whose dimension is characterized by the degrees of freedom associated with each type of disorder. For example, variations of the static gate voltages on each of the three islands (here translated to effective Cooper pair number) define a three-dimensional landscape characterizing performance for variations from the ideal offset charge. To showcase this landscape, we consider a two-dimensional slice that examines the variations in two of the three charge noise parameters, shown in Fig.~\ref{fig:A6}. A landscape of this type exists for flux noise (where it is a three-dimensional space) as well as for fabrication disorder (where it is a six-dimensional space). 

In what follows, rather than attempting to tune a given device to a perfect operating point, we instead consider a statistical analysis that samples from all dimensions of each landscape to find the fraction of devices with low insertion loss, and of those devices, the distribution of their minimum extinction ratios, specifically highlighting a median device outcome. This would correspond, in the case of static disorder, to finding high performing devices, and provides a yield of viable devices given a performance metric.

\subsection{Statistical Analysis}
\label{app:SA}

\begin{figure}
\centering
\includegraphics[width = 0.48 \textwidth]{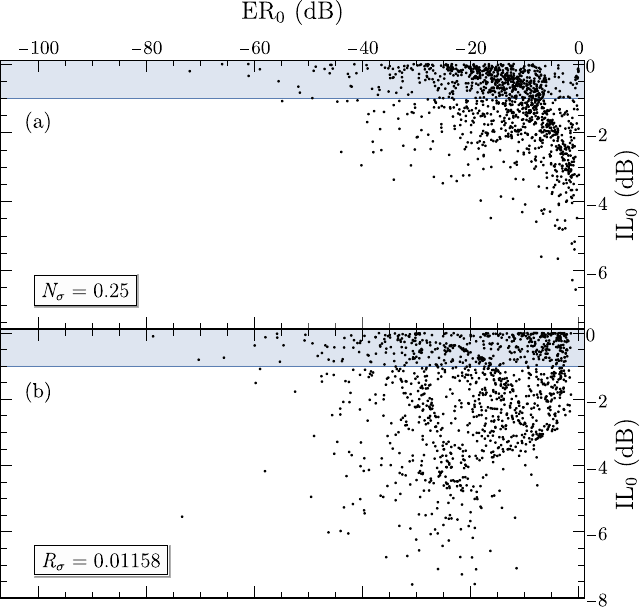}
\caption{A pairwise plot of the minimum of the extinction ratio $\text{ER}_{\text{0}}$ and the maximum of the insertion loss $\text{IL}_{\text{0}}$ in the presence of (a) charge noise for 1200 samples from a noise distribution with a mean of $N_\mu = 0$ and a standard deviation $N_\sigma = 0.25$ and (b) flux noise for 1200 samples from a noise distribution with a mean of $R_\mu = 0$ and a standard deviation $R_\sigma = 0.01158$. Shading indicates the device yield $\Upsilon$--- the fraction of points that constitute $\text{ER}^\Upsilon_{\text{0}}$. Parameters and offset charge for ideal tuning are the same as in Fig.~\ref{fig:PDplot}. Results shown are specifically for transmission between ports one and two, but are representative of the circuit overall due to circuit symmetry and the sampling from identical noise distributions at each island.}
\label{fig:A3}
\end{figure}

Upon introducing either charge noise or flux noise (as described in the main text), sampled from a normal distribution with standard deviation $N_\sigma$ or $R_\sigma$, respectively, we examine both $\text{ER}_{\text{0}}$ and $\text{IL}_{\text{0}}$ for a total of 1200 noise samples. As an example, Fig.~\ref{fig:A3}a shows the pairwise plot of $\text{ER}_{\text{0}}$ and $\text{IL}_{\text{0}}$ for 1200 charge noise samples for which $N_\sigma=0.25$. Similarly, Fig.~\ref{fig:A3}b shows the same pairwise plot but for 1200 flux noise samples for which $R_\sigma \approx 0.012$.

We then impose a performance threshold, examining only those noise samples for which there is less than 1 dB of insertion loss ($\text{IL} \geq -1$), shown visually by the points within the shaded regions of Fig.~\ref{fig:A3}. We define this subset of points as $\text{ER}^\Upsilon_{\text{0}}$, characterizing this distribution by calculating its quartiles for increasing $N_\sigma$ or $R_\sigma$. Figs.~\ref{fig:num}c and \ref{fig:num}d of the main text display the median of this distribution $\widetilde{\text{ER}}^\Upsilon_{\text{0}}$ with error bars from the first to third quartiles as a function of $N_\sigma$ and $R_\sigma$, respectively.

We also define the device yield $\Upsilon$ as the probability for transmission to exceed the imposed threshold ($\text{IL} \geq -1$). This is approximated by
\begin{equation}
\Upsilon = \frac{\text{\# of Samples with IL}\geq -1}{\text{Total \# of Samples}} \enspace .
\end{equation}
Figs.~\ref{fig:num}e and \ref{fig:num}f display this probability for increasing $N_\sigma$ and $R_\sigma$, respectively. Note that this procedure is also utilized for the fabrication disorder introduced and described below in Appendix~\ref{app:FD}.

\subsection{Fabrication Disorder}
\label{app:FD}

In addition to considering charge and flux disorder, we also examine the effects of fabrication disorder in the individual Josephson junctions of the model, shown in Fig.~\ref{fig:schematic}b. The high symmetry present in our analysis also extends to the junctions--- we assume identical Josephson energies $E_\text{J}$ and junction capacitances $C_\text{J}$ (yielding identical charging energies $E_\text{C} = \frac{e^2}{2 C_\text{J}}$) for all six junctions. By separately introducing disorder in these quantities, we can then assess the model's subsequent performance when subject to variations in junction parameters. 

\begin{figure}[H]
\centering
\includegraphics[width = 0.48 \textwidth]{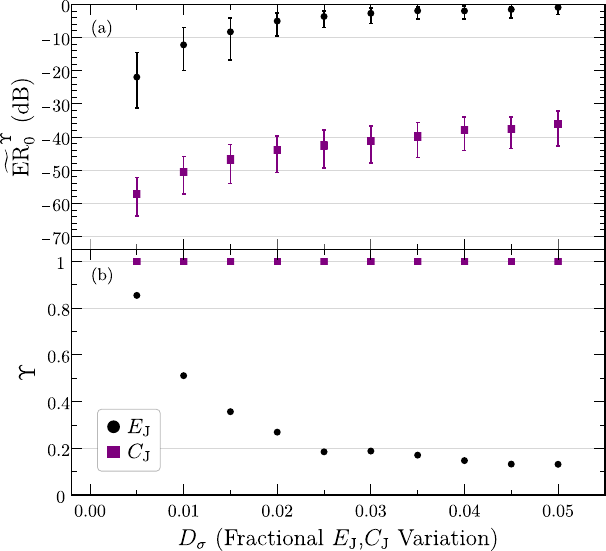}
\caption{(a) The median of $\text{ER}^\Upsilon_\text{0}$ (defined in Appendix~\ref{app:SA} as the extinction ratio for devices with less than 1 dB of insertion loss), $\widetilde{\text{ER}}^\Upsilon_\text{0}$, is plotted as a function of the standard deviation of the $E_\text{J}$ or $C_\text{J}$ noise distribution $D_\sigma$ as indicated by the legend. Error bars indicate the spread from the first to third quartile of $\text{ER}_\text{0}^\Upsilon$ for each value of $D_\sigma$. (b) Device yield $\Upsilon$ (defined in Appendix~\ref{app:SA} as the fraction of devices with less than 1 dB of insertion loss) is plotted as a function of $D_\sigma$. Parameters for ideal tuning are the same as in Fig.~\ref{fig:num}b.}
\label{fig:newnoise}
\end{figure}

To test the system's fabrication tolerances, we begin with the system tuned for ideal circulator behavior with all six Josephson junctions having identical $E_\text{J}$ and $C_\text{J}$. Analogous to the charge and flux noise cases, we then introduce a variation $\Delta''$ in either $E_\text{J}$ or $C_\text{J}$ that is unique for each junction. In both the $E_\text{J}$ and $C_\text{J}$ noise cases, we take $\Delta''$ to be a fraction of the original value, randomly sampling from a normal distribution with a mean of $D_\mu = 0$ and a standard deviation $D_\sigma$. As detailed and defined above in Appendix~\ref{app:SA}, we assess performance in the presence of either $E_\text{J}$ or $C_\text{J}$ noise by examining $\text{ER}^\Upsilon_\text{0}$ and $\Upsilon$ as a function of $D_\sigma$, shown together in Figs.~\ref{fig:newnoise}a and \ref{fig:newnoise}b, respectively. As stated above, $\text{ER}^\Upsilon_\text{0}$ and $\Upsilon$ convey the distribution of the minimum of the extinction ratio for those noise samples resulting in in less than 1 dB of insertion loss and the likelihood that this is the case using a sample size of 1200 noise samples.

Upon increasing $D_\sigma$, we observe markedly different performance between the introduction of noise in $E_\text{J}$ versus that in $C_\text{J}$, as shown in Fig.~\ref{fig:newnoise}. In particular, we see that while the introduction of noise in the Josephson energies of the junctions degrades performance, varying the junction capacitances does not. This is best understood by considering that the original ideal parameters are such that $E_\text{J} = 4 E_\text{C} \approx 40 E_\Sigma$, thereby necessitating that any variations introduced in $E_\text{J}$ are bound to have a larger effect than those in $C_\text{J}$ (and subsequently $E_\text{C}$ and $E_\Sigma$). Nonetheless, Fig.~\ref{fig:newnoise} would indicate that variations of approximately 1.5 \% in $E_\text{J}$ may be tolerated. Beyond 2 \%, nonreciprocity is extinguished, where we see that $\widetilde{\text{ER}}^\Upsilon_{\text{0}} \approx 0$ and $\Upsilon \lesssim$ 20 \%. In contrast, Fig.~\ref{fig:newnoise} shows that variations up to 5 \% (and likely higher) in $C_\text{J}$ can be tolerated--- for all distribution sizes $\widetilde{\text{ER}}^\Upsilon_{\text{0}} < -30$ and $\Upsilon =$ 100 \%. 

\end{document}